\documentclass[useAMS,usenatbib]{mn2e}
\usepackage[cp1251]{inputenc} 
\usepackage{graphicx}
\usepackage{amssymb}
\usepackage[english]{babel}
\usepackage{captcont}
\usepackage{morefloats}
\usepackage{datetime}

\usepackage[dvipsnames]{color}

\def\ltsima{$\; \buildrel < \over \sim \;$}
\def\simlt{\lower.5ex\hbox{\ltsima}}
\def\gtsima{$\; \buildrel > \over \sim \;$}
\def\simgt{\lower.5ex\hbox{\gtsima}}

\newcommand{\kmpskpc}{km~s\ensuremath{^{-1}}~kpc\ensuremath{^{-1}}}
\newcommand{\kmps}{km s\ensuremath{^{-1} }}
\newcommand{\Msun}{M\ensuremath{_\odot}}
\newcommand{\Oo}{\displaystyle}

\title[Spirals beyond the optical radius]{Spiral density waves in the outer galactic gaseous discs}

\pagerange{\pageref{firstpage}--\pageref{lastpage}} \pubyear{2015}

\def\LaTeX{L\kern-.36em\raise.3ex\hbox{a}\kern-.15em
    T\kern-.1667em\lower.7ex\hbox{E}\kern-.125emX}

\author[S. Khoperskov and G. Bertin]{ S.A. Khoperskov$^{1,2,3}$\thanks{Sergey.Khoperskov@unimi.it} and G. Bertin$^{1}$ \\ $^1$Dipartimento di Fisica, Universit\`{a} degli Studi di Milano, via Celoria 16, I-20133 Milano, Italy \\ $^2$Institute of Astronomy, Russian Academy of Sciences, Pyatnitskaya st., 48, 119017 Moscow, Russia\\ $^3$Sternberg Astronomical Institute, Moscow M.V. Lomonosov State University, Universitetskij pr., 13, 119992 Moscow, Russia}

\begin{document}
    
\maketitle

\begin{abstract}
Deep HI observations of the outer parts of disc galaxies demonstrate the frequent presence of extended, well-developed spiral arms far beyond the optical radius. To understand the nature and the origin of such outer spiral structure, we investigate the propagation in the outer gaseous disc of large-scale spiral waves excited in the bright optical disc. Using hydrodynamical simulations, we show that non-axisymmetric density waves, penetrating in the gas through the outer Lindblad resonance, can exhibit relatively regular spiral structures outside the bright optical stellar disc. For low-amplitude structures, the results of numerical simulations match the predictions of a simple WKB linear theory. The amplitude of spiral structure increases rapidly with radius. Beyond $\approx 2$ optical radii, spirals become nonlinear (the linear theory becomes quantitatively and qualitatively inadequate) and unstable to Kelvin-Helmholtz instability. In numerical simulations, in models for which gas is available very far out, spiral arms can extend out to 25 disc scale-lengths. A comparison between the properties of the models we have investigated and the observed properties of individual galaxies may shed light into the problem of the amount and distribution of dark matter in the outer halo.
\end{abstract}


\section{Introduction}
For several galaxies, deep HI images demonstrate the presence of large-scale spiral arms in extended gaseous discs well outside the bright stellar component~\cite[e.g., see][]{2008A&ARv..15..189S}. The most spectacular cases are~NGC~1512/1510, NGC~5055, NGC~6744, NGC~6946, and NGC~5236.  Some spirals exhibit a rather complex morphology~\cite[e.g., NGC~6946;][]{2008A&A...490..555B}; yet, symmetric grand-design structures are not so rare~\cite[e.g., NGC~1512;][]{2009MNRAS.400.1749K}. 

UV images of the extended galactic (XUV) gaseous disc point to ongoing star formation even in such outermost regions~\citep{2008AJ....136.2846B}. In particular, GALEX data provide evidence for weak but significant star formation in such very-low gas-density environments. HII regions have been noted far away, outside the bright optical disc~\citep{2011MNRAS.415L..31K}. In addition, the gaseous layer in the disc plane is characterized by a rather high velocity dispersion $1-10$~\kmps. An interpretation of these findings is required. 

 A comparison of HI and UV brightness of the extended outer disk shows that spiral arms are a necessary, but not sufficient, requirement for star formation~\citep{2012ApJ...757...64B}. Interestingly, although with only little attention paid to the dynamical origin of spiral structure, the simulations by~\citet{2008ApJ...683L..13B}~\citep[see also][]{2010ApJ...713L..780B} show that organized compression regions and filamentary structures should occur frequently, at least if the extended gas is characterized by a constant surface density at the level $5-10$~\Msun~pc$^{-2}$ (which is on the high side, with respect to HI observations; note also that this implies a much shorter depletion time scale than expected~\citep{2010AJ....140.1194B}). However, in spite of the fact that there are many examples of extended UV discs, we still lack firm theoretical models of star formation outside the optical radius in external galaxies. This difficulty is largely related to our poor understanding of the physical conditions in those regions, both in terms of general properties of the interstellar medium and of the resulting stellar populations   ~\citep{2012ApJ...749...20K,2013ApJ...775...40B}. It appears that, much like in some dwarf galaxies, gas metallicity and density are low~\citep{2007ApJ...661..115G}. On the other hand, judging from the observed UV emission, star formation might be expected to be supported for at least a few billion years. Yet the efficiency of transforming gas into a young stellar population is expected to be  low~\citep{2010AJ....140.1194B}, with a time scale $\approx 50$ times longer than in normal disks. Note that cluster counts in the outskirts of M~83 appear to be consistent with model predictions based on a standard IMF and cluster aging effects; therefore even low-mass newborn stellar clusters ($10^{2-3}$\Msun) occasionally have O stars~\citep{2012ApJ...749...20K}.

The thickness of the disc is likely to increase with radius. A proper model of this feature and of the dynamical mechanisms involved in star formation and spiral structure noted above would also depend on the characteristics (amount and spatial distribution) of the dark matter present, which are not easy to measure and are obviously of great astrophysical interest. 

Thermal instabilities are probably less important, because of the low metallicity of the gas. Thus, under these conditions, to produce star-forming regions a mechanism is needed able to compress the gas on scales of about one kiloparsec. This suggests that a hierarchy of perturbations should be operating, allowing gas to form stars at small scales. The perturbation on the largest scale is the one associated with grand-design spiral patterns.

On the one hand, natural mechanisms for the generation of structures beyond the optical radius are those related to gas accretion~\citep{2009ApJ...703..785D}, streams from gas-rich companions in close passages~\citep{2010MNRAS.403..625D}, and, in general, tidal interactions with satellites or dwarf galaxies~\citep{2005ApJ...635..931B}. In these scenarios, the role of dark matter is not clear. On the other hand, it is generally believed that, in the presence of a triaxial (i.e., non-axisymmetric) distribution of dark matter~\citep{2012ARep...56...16K,2014AJ....147...27V}, disc-halo interactions could generate spiral structure. However, for the majority of galaxies our knowledge of the three-dimensional distribution of dark matter is quite poor. 

Hydrogen is mostly neutral in the outermost regions, so that the relatively low photoionization level expected makes it likely that large-scale magnetic fields should not play an important role in structure and star formation. Although it is generally believed that gravitational instabilities are the main driver for spiral structure formation, the collective processes occurring in the outer gaseous medium of galaxies remain largely unexplored.

The study of the global large-scale spiral structure has a long history. The commonly accepted picture is that spiral structure is the manifestation of density waves in the galactic disc~\citep{1964ApJ...140..646L}, for which gas and stars cooperate collectively~\citep{1966PNAS...55..229L}.  Currently, the picture that the grand-design spiral patterns are associated with few self-excited global spiral modes is the scenario that has been worked out in greatest quantitative detail~\cite[see][and references therein]{1996ssgd.book.....B}, supported by a number of successful observational tests. Convincing tests from realistic numerical simulations remain difficult to obtain, because of the complexity of the physical phenomena involved, ranging from the role of resonances in the collisionless stellar component to the destabilizing and self-regulating role of the dissipative cold interstellar medium~\citep{1993A&A...272...37E,2009ApJ...706..471B,2011ApJ...730..109F}.

In this paper we extend the study of spiral patterns outside the optically-bright stellar disc presented by \citet{2010A&A...512A..17B}. We perform 3D~hydrodynamical non-linear simulations in simple galaxy models outside a central disc; we assume that in the central disc spiral structure is dominated by one or few modes, which act as a central ``engine" for what is observed in the outer parts. We then study the properties of spiral structure in the outer gaseous disc by varying a set of parameters that characterize the relevant perturbations. If the conditions in the galactic disc favour the leakage of small-amplitude ($\leq 10\%$) non-axisymmetric density waves through the outer Lindblad resonance (OLR), we find that indeed these perturbations can give rise to large-scale prominent spirals covering a wide radial range beyond the optical radius of the galaxy.  Note that outside OLR the gas can support density-wave propagation even when its effective velocity dispersion is above the condition of marginal stability with respect to axisymmetric perturbations (or, correspondingly, the column density is below the related critical density).

The paper is organized as follows. Section 2 describes the adopted basic model and numerical approach. In Sect.~3, we present the main results for various models. In particular, we address models in which the inner disc is dominated by a single mode and models with two or three important modes, incorporated as a boundary condition at an inner annular region in the outer parts of the stellar disc. The impact of small-scales inhomogeneities and subgrid physics are also discussed briefly. Our nonlinear simulations are compared with the results of the linear theory. Section 4 provides discussion and conclusions.

\section{Model}

\begin{figure}
\includegraphics[width=1\hsize]{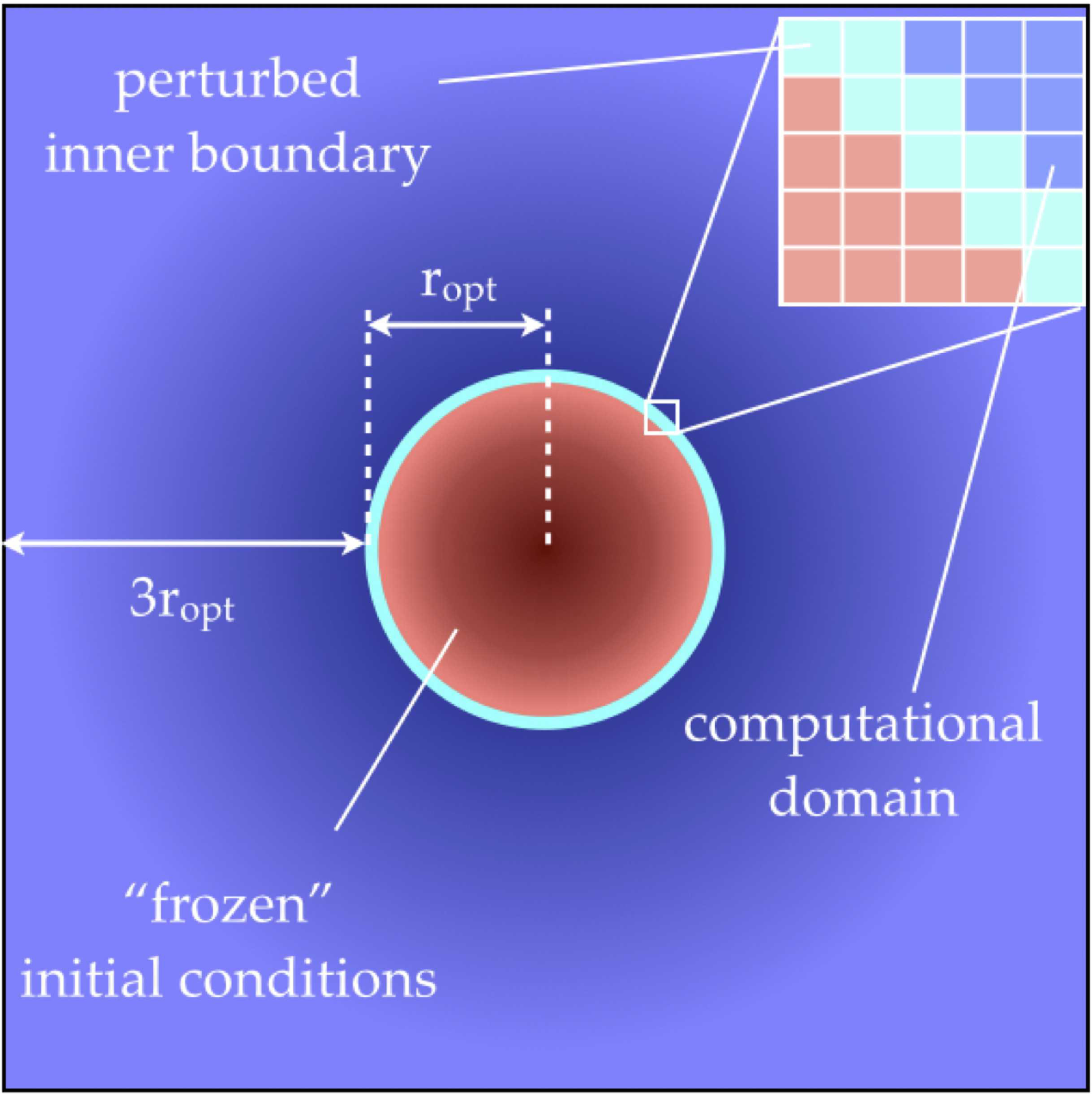}
\caption{General scheme of the computational model. A fixed area of initial conditions inside the optical radius, defined here as $r_{\rm opt}=6h$, is shown as a red disc. The thin, light circular annulus (cyan line) is the site where the perturbation is imposed, which is the inner boundary for the evolving domain. As shown on the top right corner, the cyan line corresponds to the two computational grid cells where the initial conditions are designed 
to mimic the presence of an outgoing density wave signal coming from the inner regions where coherent spiral structure is present. The blue area outside the red disc is the actual computational domain from $6h$ out to $24h$, where the hydrodynamical quantities evolve.}\label{fig::scheme} 
\end{figure}

The pure hydrodynamical approach based on the TVD MUSCL~(Total Variation Diminishing Multi Upstream Scheme for Conservation Laws) scheme is used for numerical simulations. The computational technique was described by~\cite{2014JPhCS.510a2011K}. Here we present a set of 3D~simulations on a uniform $2048\times2048\times128$ Cartesian grid. For a fiducial model, the computational box is $144\times144\times9$~kpc with a spatial resolution of about $70$~pc. A study of resolution effects has been performed with cell size of $200$ and $35$~pc~(see~Sect.~\ref{sec::res_study}).  The general set up of the simulations is illustrated in Fig.~\ref{fig::scheme}.

\subsection{Basic state}

Below we deal with the evolution of a gaseous disc embedded in the fixed external potential of an approximately isothermal dark matter halo $\Phi_h$~\cite[model introduced by][]{1995ApJ...447L..25B}  combined with the potential of a stellar disc $\Phi_d$ of the Miyamoto-Nagai form~\citep{1975PASJ...27..533M}. In dimensional units we assume that $h = 3$~kpc is the stellar disc exponential scale length. 

\begin{figure}
\includegraphics[width=1\hsize]{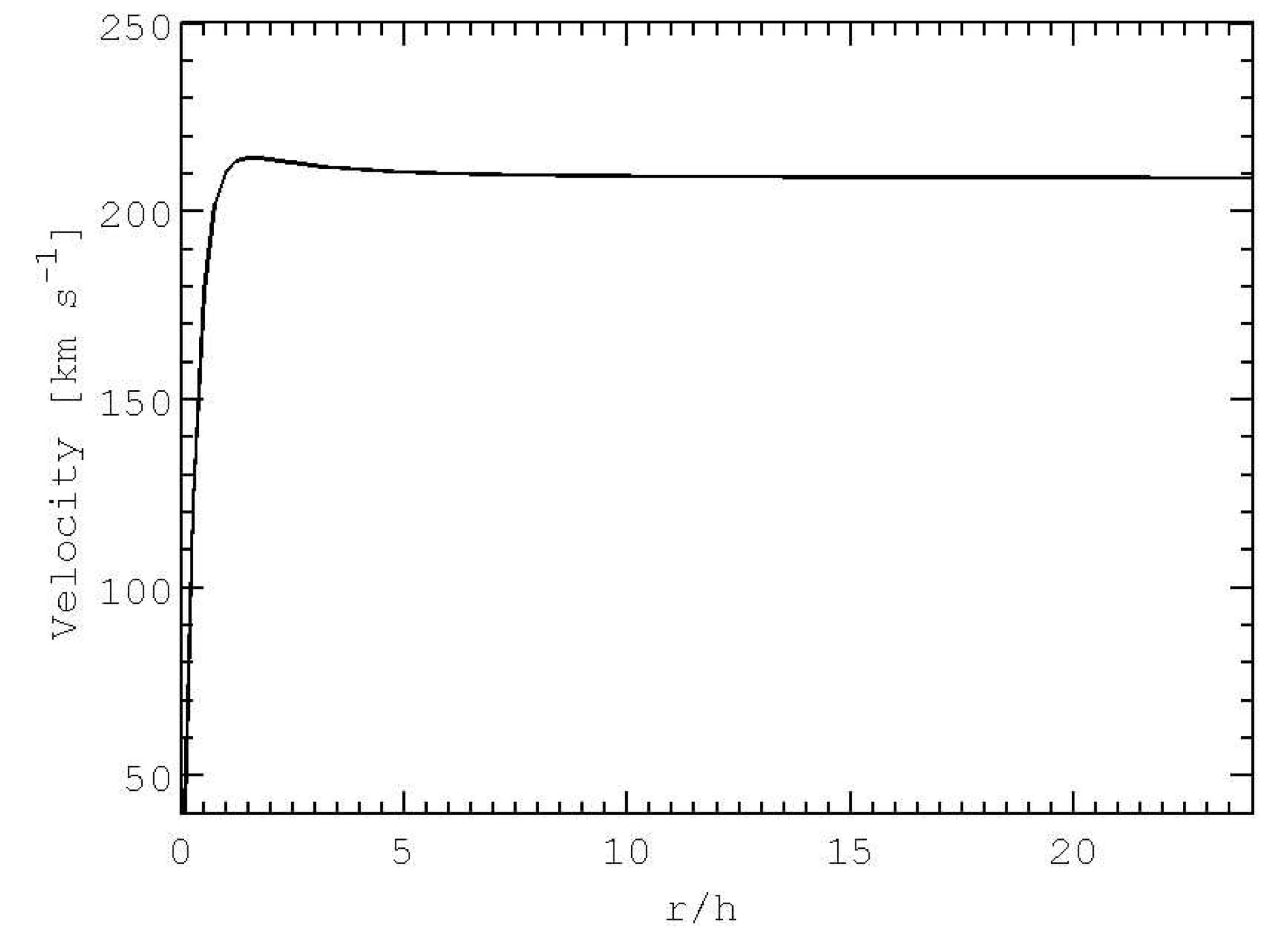}
\caption{Adopted rotation curve $V(r)$.}\label{fig::rotation curve} 
\end{figure}

\begin{figure}
\includegraphics[width=1\hsize]{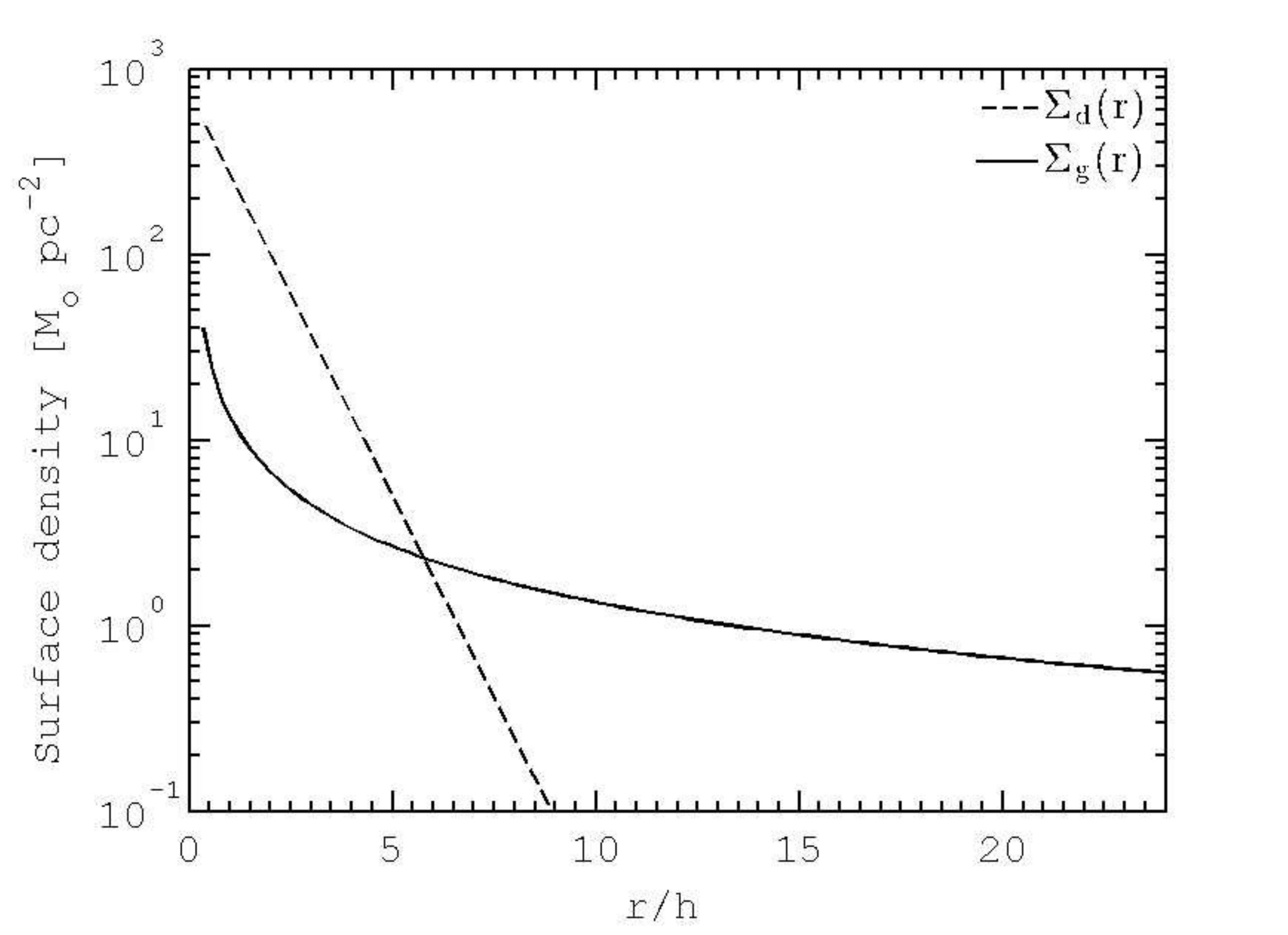}
\caption{Initial surface density distributions of gas (solid line) and stars (dashed line).}\label{fig::surfacedensity}
\end{figure}

The parameters of the external potential were chosen to support the flat rotation curve of the gaseous disc~(see~Fig.~\ref{fig::rotation curve}).  For the majority of nearby galaxies,  the gas distribution follows an exponential profile within the optical radius $r_{\rm opt}$~\citep{2012ApJ...756..183B}. At far away distances, gas density profiles are not so well known. Nevertheless,  it is natural to assume a $\Sigma_{\rm g} \propto 1/r$ profile of gas beyond $(1-2)r_{\rm opt}$. For simplicity, we adopt as initial gas density distribution the $\Sigma_{\rm g} \propto 1/r$ profile also for the inner part of the disc, which is kept fixed in our simulations. The initial  stellar disc surface density and gaseous surface density profiles are shown in Fig.~\ref{fig::surfacedensity}.

For the value of the gas velocity dispersion we adopt that obtained from the marginal-stability condition for a non-self-gravitating layer of finite thickness, as is expected in the galactic outskirts~\cite[see~Eqs.~(15.21), (15.22), and Fig.~15.4 in][]{2014dyga.book.....B}: 
\begin{equation}\label{eq::QteqQ}
Q = Q_{\rm max} = 0.425\,.
\end{equation}
Thus, using the epicyclic frequency $\kappa$ and $\Sigma_g$ from our initial conditions we have the relation for the radial velocity dispersion:
\begin{equation}\label{eq::cr}
 c = 0.425 \pi G \Sigma_g / \kappa\,,
\end{equation}
which, outside the circle $r = 6 h$, is approximately constant $c\approx 3.4$~\kmps. Such value is consistent with the observational data, which suggest a velocity dispersion of the gas clouds in the range $1-10$~\kmps~\citep{1989ApJ...339..763S,2004ARA&A..42..211E}.

\begin{figure}
\includegraphics[width=1\hsize]{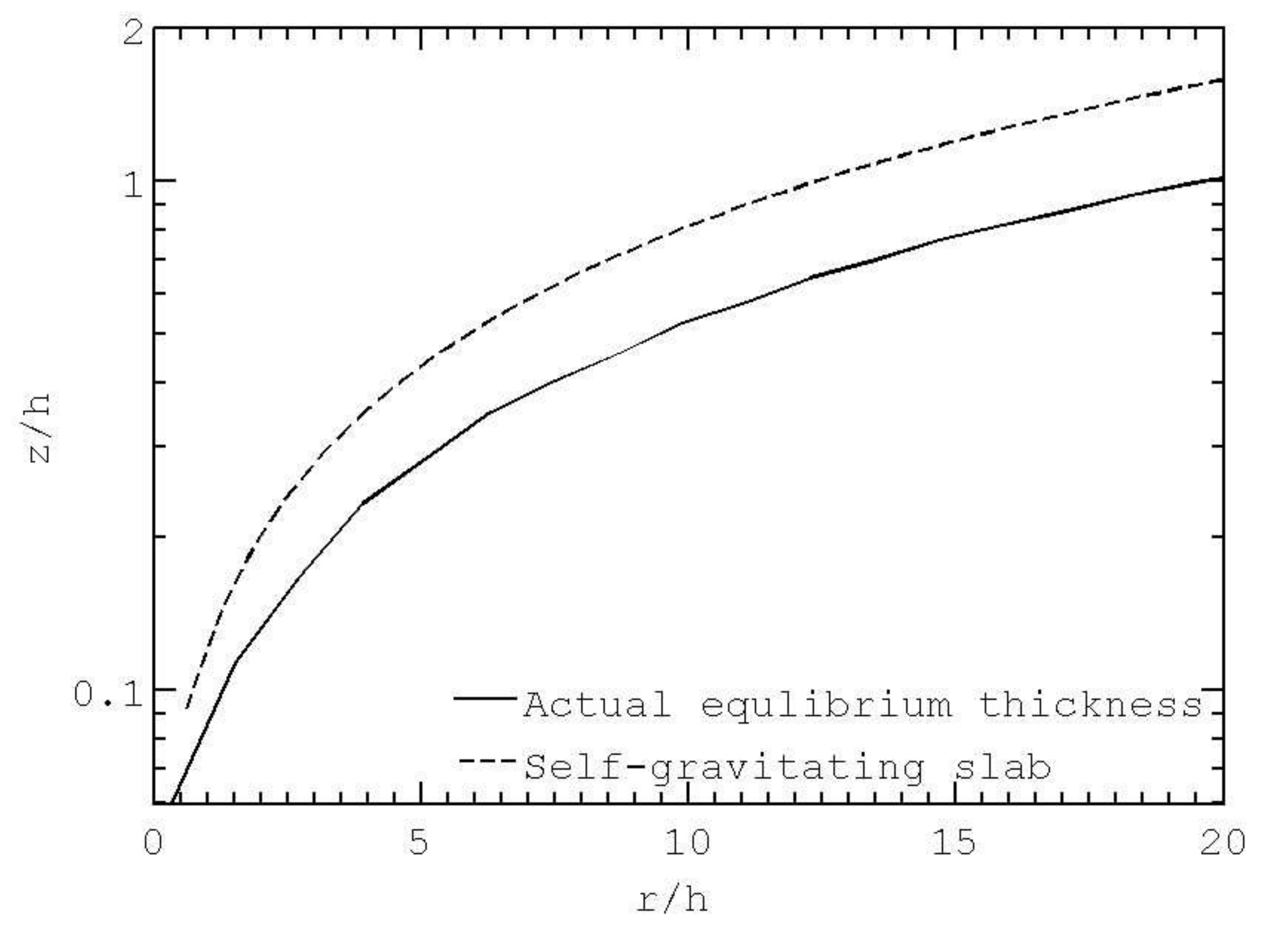}
\caption{The solid line is the gaseous disc scale height obtained from the vertical equilibrium according to Eq.~(\ref{eq::vert_equlibrium}). The dashed line is the disc thickness according to the marginal-stability condition of a fully self-gravitating slab only~(Eq.~\ref{eq::z0_margst}).}\label{fig::thickness} 
\end{figure}

The gaseous disc thickness is set by the condition of vertical hydrostatic equilibrium:
\begin{equation}\label{eq::vert_equlibrium}
\Oo \frac{1}{\rho}\frac{\partial p}{\partial z} + \frac{\partial\Phi}{\partial z} = 0\,,
\end{equation}
where the gas volume density $\rho$ and pressure $p$ are connected by the equation of state $p = \rho c^2$ and the total gravitational potential $\Phi = \Phi_h + \Phi_d + \Phi_g$ takes into account the potential of the gas $\Phi_g$, which is the solution of the Poisson equation:
\begin{equation}\label{eq::poisson}
\Oo \frac{1}{r}\frac{\partial}{\partial r} \left( r \frac{\partial\Phi_g}{\partial r} \right) + \frac{\partial^2 \Phi_g}{\partial z^2} = 4\pi G \rho\,.
\end{equation}
Eqs~(\ref{eq::vert_equlibrium}) - (\ref{eq::poisson}) determine the equilibrium vertical distribution of the gas. We define its vertical scale height $z_0(r)$ by minimizing the difference $F(z_0)$:
\begin{equation}
F(z_0) = |\rho_0 \cosh^{-2}(z/z_0)-\rho(r,z)|\,.
\end{equation}  
As the vertical gravity decreases with radius then the equilibrium disc thickness increases. Figure~\ref{fig::thickness} shows such disc flaring from the solution of Eqs.~(\ref{eq::vert_equlibrium}) - (\ref{eq::poisson}) for our initial parameters. We also compare our solution with the gas thickness profile supported by the gas self-gravity alone~\cite[see~Eq.~(14.12) in][]{2014dyga.book.....B}:
\begin{equation}
\Oo \frac{z_0}{r} = 0.18 \frac{\pi G \Sigma_g}{r \kappa^2}\,,\label{eq::z0_margst}
\end{equation}
where the coefficient $0.18$ is obtained from the marginal stability condition~\cite[see Fig.~2 in][]{2010A&A...512A..17B}.

\subsection{The imposed perturbations}

\begin{figure}
\includegraphics[width=1\hsize]{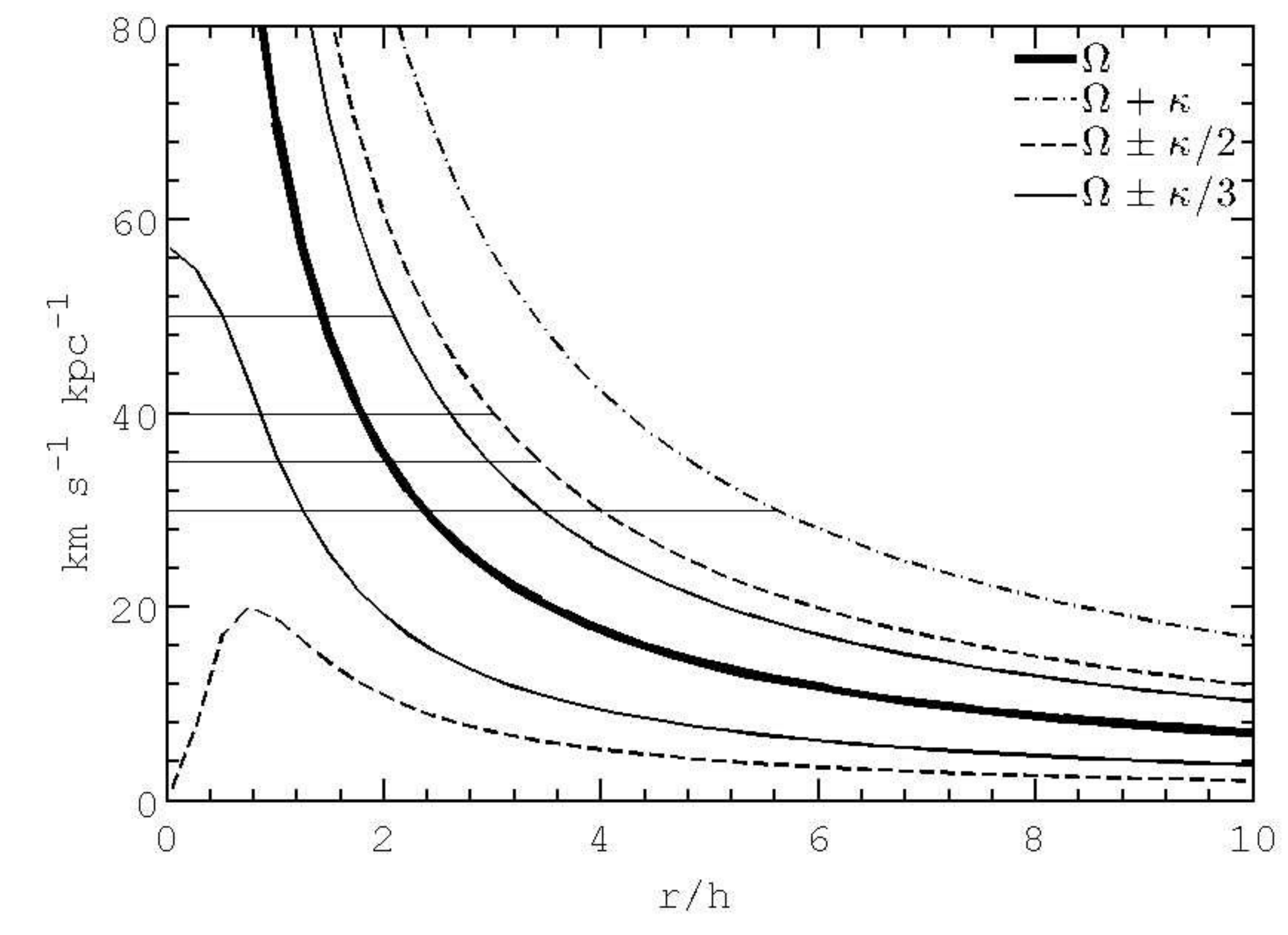}
\caption{Kinematics of the gaseous disc. The thick solid line is the angular velocity $\Oo \Omega = V(r)/r$, the thin solid lines represent $\Omega \pm \kappa/3$,  the dashed lines $\Omega \pm \kappa/2$ and the dash-dotted line is $\Omega + \kappa$. The horizontal thin solid lines correspond to the pattern speeds of the modes considered in our simulations: from top to bottom, 50~\kmpskpc (for the $m=3$ mode),  40~\kmpskpc (for one $m=2$ mode),  35~\kmpskpc (for a second $m=2$ mode),  30~\kmpskpc (for the $m=1$ mode). The mode parameters are summarized in Table~\ref{tab::ini}. All the selected modes have OLR inside the inner boundary of the computational domain~(see also Fig.~\ref{fig::scheme}).}\label{fig::diagramm} 
\end{figure}

We now introduce the properties of the perturbation imposed  at the inner boundary~(see Fig.~\ref{fig::scheme}). Basically, we consider the perturbations of the hydrodynamical quantities ${\bf \hat{X}}$ at the inner boundary~(cyan thin circular annulus in Fig.~\ref{fig::scheme}) to be proportional to $ \cos\left(m\theta- \omega t\right)\,,$ where  $m$ is the mode azimuthal number, $\Omega_p=\omega/m$ is the angular speed of the spiral density perturbation, $t$ is the time, $\theta$ is the angular coordinate. We assume that the relative amplitude of the density wave is equal to $A_0$~(mean values are shown in Table~\ref{tab::ini}) and then relative amplitudes of the perturbation for all other quantities can be found straightforwardly from Eqs. (7), (8), (9), and (13) for the short-trailing wave-branch in~\cite{2010A&A...512A..17B}. In the calculation, the perturbation at the inner boundary is applied only for the hydrodynamical evolution of the outer disc. Thus the gas distribution is kept ``frozen" and axisymmetric inside the disc defined by the circle $r_{\rm opt}=6h$. For any given mode considered in our study, the value of $\Omega_p$ sets an outer Lindblad resonance inside the inner boundary at $r_{\rm opt}=6h$. Thus the perturbations that we consider propagate outward~(see~Fig.~\ref{fig::diagramm}).

It is believed that several spiral modes generally coexist in a given disc model~\citep{1977PNAS...74.4726B, 2000ApJ...541..565K}. Then we consider not only a model where a single dominant mode is present, but also models that include a superposition of more than one mode. In this case the  density perturbations can be written in the following form:
\begin{equation}\label{eq::pertr2}
\Oo  \Sigma_1 = \sum_i A_{0,i}\cos\left(m_i\theta - \omega_{i} t + \delta_i \right)\,,
\end{equation}
where $A_{0,i}$ are the relative amplitudes of the density perturbation at the inner boundary, $\delta_i$ are the initial phases of perturbations and $\omega_i = \Omega_{p,i}/m_i$. For the simplicity, in the following we take $\delta_i=0$. Much like for the case of a single mode perturbation model, amplitudes for pressure and velocities are calculated according to the expressions provided by~\cite{2010A&A...512A..17B}. We consider three models with multi-mode boundary conditions (see detailed parameters in Table~\ref{tab::ini}).  
In particular, we take a case that includes the superposition of a one-armed and a two-armed
perturbation~(model H1), a case with a pair of two-armed patterns~(model F1), and the case made of three modes, each with a different number of arms~(model J1). The amplitudes and the corresponding pattern speeds of the perturbations for these models are chosen so as to be in qualitative agreement with the linear theory of global spiral modes. 
Of course, other combinations of perturbation parameters could be reasonable: our models should only be considered as a simple 
representation of typical cases broadly consistent with the modal density wave theory~\cite[see][]{1977PNAS...74.4726B,1996ssgd.book.....B}. 

\begin{table*}
\begin{center}
\caption{Parameters of different runs, where $m$ is the perturbation number of arms, $A_0$ is the relative amplitude of the adopted density perturbation, $\Omega_p$ is the pattern speed, $ r_{\rm co} $ is the corotation radius, $r_{\rm OLR}$ is the outer Lindblad resonance position, $h=3$~kpc is the exponential scale length of the stellar disc.  When several parameters are present in a given column the inner boundary is perturbed with more than one mode. B1 is the reference model.}\label{tab::ini}
\begin{tabular}{ccccccc}
\hline
Run   & $m$ & $A_0$ & $\Omega_p$ & $ r_{\rm co} $ & $ r_{\rm OLR} $ & Additional  \\
   &         &              &  \kmpskpc     & 		$h$					   & 		$h$					&  feature \\
\hline
E1       & 1    	 	 	& 0.05    					 & 30   		 			& 2.4 & 5.6 & -     			\\ 
B1      & 2    	 		& 0.1     					 & 40  		 			& 1.8 & 3 	& -  		   		\\ 
B2       & 2    	 		& 0.1      					 & 40   		 			& 1.8 & 3 	& clumpy gas distribution   		  	\\ 
B4       & 2    	 		& 0.1      					 & 40   		 			& 1.8 & 3 	& potential perturbation    \\ 
B5       & 2    	 		& 0.1      					 & 40   		 			& 1.8 & 3 	& subgrid cooling     	\\ 
B7       & 2    	 		& 0.1      					 & 40   		 			& 1.8 & 3 	&  higher velocity dispersion $c = 5$ \kmps   		  	\\ 
K1       & 3    	 		& 0.15      				 & 50   		 			& 1.4 & 2.1  & -     	\\ 
H1       & 1 / 2 		    & 0.05 / 0.1  			 & 30 / 40   			& 2.4/1.8& 5.6 / 3 & -     			\\ 
F1       & 2 / 2   		& 0.07 / 0.1   			 & 35 / 40  			& 2/1.8 & 3.4 / 3 & -     			\\ 
J1       & 1 / 2 / 3    & 0.05 / 0.1 / 0.15  & 30 / 40 / 50   & 2.4/1.8/1.4& 5.6 / 3 / 2.1& -     			\\ 
\hline
\end{tabular}
\end{center}
\end{table*}

To avoid an initial exaggerated kick on the disc, we let the perturbation amplitudes grow from vanishingly small to finite values according to a linear law during one typical rotation period $T_1$ until they reach the chosen value $A_0$. From Fig.~\ref{fig::rotation curve}, our time scale is the rotation period $T_1 \approx 0.5$~Gyr at $6h = 18$~kpc.

Initially we set up the dynamical equilibrium of the gaseous disc within the entire computational domain, that is, in both the red and the blue areas of Fig.~\ref{fig::scheme}. Then we ignore the complex self-consistent evolution of the stellar-gaseous disc within the optical radius. Namely, we keep the central part of the computational domain $r<r_{\rm opt}=6h$ as ``frozen"~(see Fig.~\ref{fig::scheme}). This defines an inner boundary layer at $r_{\rm opt}=6h$ for the live gaseous disc outside. 

It is believed that the rotation of the galaxy outside the bright optical disc is mainly supported by the gravitational potential of the dark matter, which in turn is expected to have significant substructures~\citep{1999ApJ...524L..19M}. In fact, numerical simulations of galaxy and structure formation in the cosmological context predict that galactic dark matter haloes should contain a population of so-called subhaloes~\citep{2008MNRAS.386.2135G,2004MNRAS.355..819G}. The relative motions of these substructures should induce local time-dependent variations of the gravitational field. This feature might have a strong impact at the periphery, where the baryonic matter density is small. To consider these effects we performed also simulations with a time-dependent inhomogeneous gravitational field.

The simulation of the dark matter dynamics at the outskirts would require high spatial and mass resolution. To avoid the related technical problems, in our numerical model we add to the initial conditions adopted for B1 a random perturbation of the halo gravitational potential. This defines model B4. That is, in the B4 model we recalculate the potential according to the rule: 
\begin{equation}
\Oo \Phi_h(r,z,t) = \Phi_h(r,z,0)\left[1 + \alpha(t) \right]\,,
\end{equation} 
where $\Oo \alpha(t)$ is a random number in the range $[-0.1; 0.1]$ which varies at each time step and is unique for cells with size $200^3$~pc$^3$.
Thus the total mass of the halo is conserved but a $10\%$ time-dependent perturbation of the gravitational potential is introduced. 

The galactic gaseous component is a cloudy medium. Within the optical disc a significant fraction of the gas mass is concentrated in giant molecular clouds. Outside the optical disc the cold gaseous phase is likely to be concentrated in neutral hydrogen clouds~(but see the picture explored by~\citet{1994A&A...285...79P}); apparently, these clouds are not forming stars in large amounts. These arguments suggest that we should consider an inhomogeneous gas distribution in our simulations. We designed model B2 so as to include a random perturbation of the gas density distribution with relative amplitude of $10\%$. The velocity field of the clouds was perturbed according to the mean velocity dispersion of about 3.4~\kmps. As was mentioned earlier,  this value appears to be realistic.

It is generally thought that the gas cloudy medium is collisional. To take into account this fact, we calculate an effective ``cooling rate" of the cloudy medium associated with inelastic collisions. We assume that in each computational cell there is a subgrid population of clouds that can lose energy as a result of collisions. Obviously the cooling rate depends on the gas density $n$ and cloud velocity dispersion at the cell. In calculations we applied the cloud collision rate based on the model by~\cite{2002MNRAS.334..684R}.  In the numerical scheme the cooling rate was used as a source term in the energy conservation equation. It is implemented with the standard technique used for the radiative cooling approximations widely adopted in simulations of galaxies and ISM.

In the next sections we consider the results of the dynamical simulations. First we discuss the reference case of a model with a single $m=2$ mode. We pay attention to the spiral morphology, its time-dependent evolution, and we check to what extent its behaviour agrees with the linear theory~(Sects.~\ref{sec::single_mode},~\ref{sec::linear}). Next we consider a more complex and realistic situation by studying the case of a clumpy gas distribution and of an inhomogeneous gravitational potential. A model with subgrid energy dissipation resulting from inelastic HI cloud collisions is also described in Sect.~\ref{sec::phys}. Finally we study cases with multi-mode perturbations~(Sect.~\ref{sec::several_modes}).

\section{Results}
\subsection{Single-mode perturbation}\label{sec::single_mode}
We now describe the results of hydrodynamical simulations of the gaseous disc evolution with a single-mode $m=2$ perturbation imposed at the inner ($r_{\rm opt}$) boundary. The values of the adopted parameters are shown in Table~\ref{tab::ini}. In the following, the total density of the gas is denoted by $\Sigma_g = \langle\Sigma_g \rangle + \Sigma_1$,  where $\langle \Sigma_g \rangle$ is the azimuth-averaged (radius-dependent) profile of $\Sigma_g$ and $\Sigma_1$ represents the spiral density wave. In general we will refer to $\Sigma_g$ in units of \Msun pc$^{-2}$; on other occasions, such as in Fig.~ 6, we will refer to the (dimensionless) relative density perturbation defined as $\Sigma_1/\langle\Sigma_g\rangle$.

\begin{figure*}
\includegraphics[width=0.24\hsize]{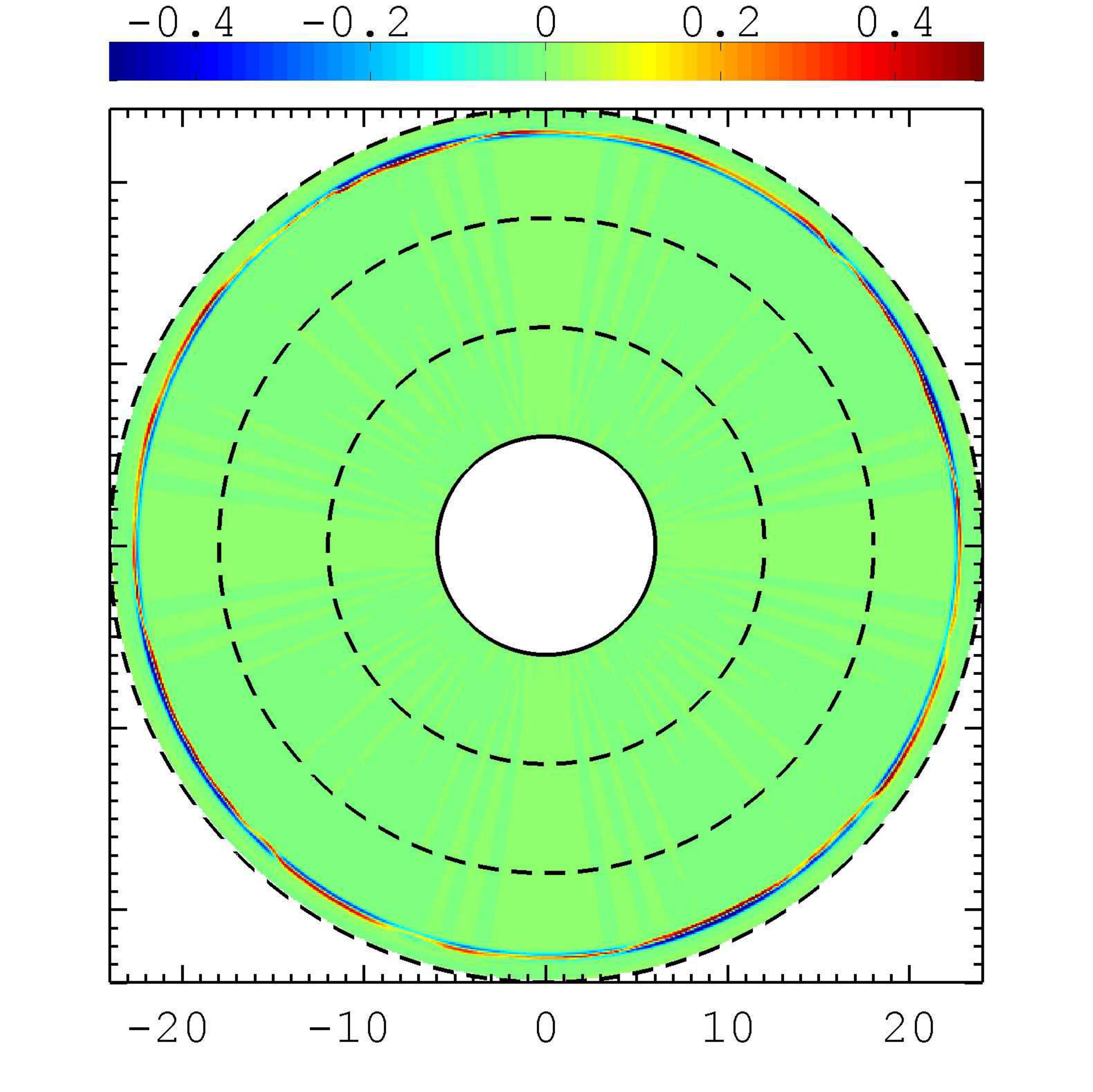}
\includegraphics[width=0.24\hsize]{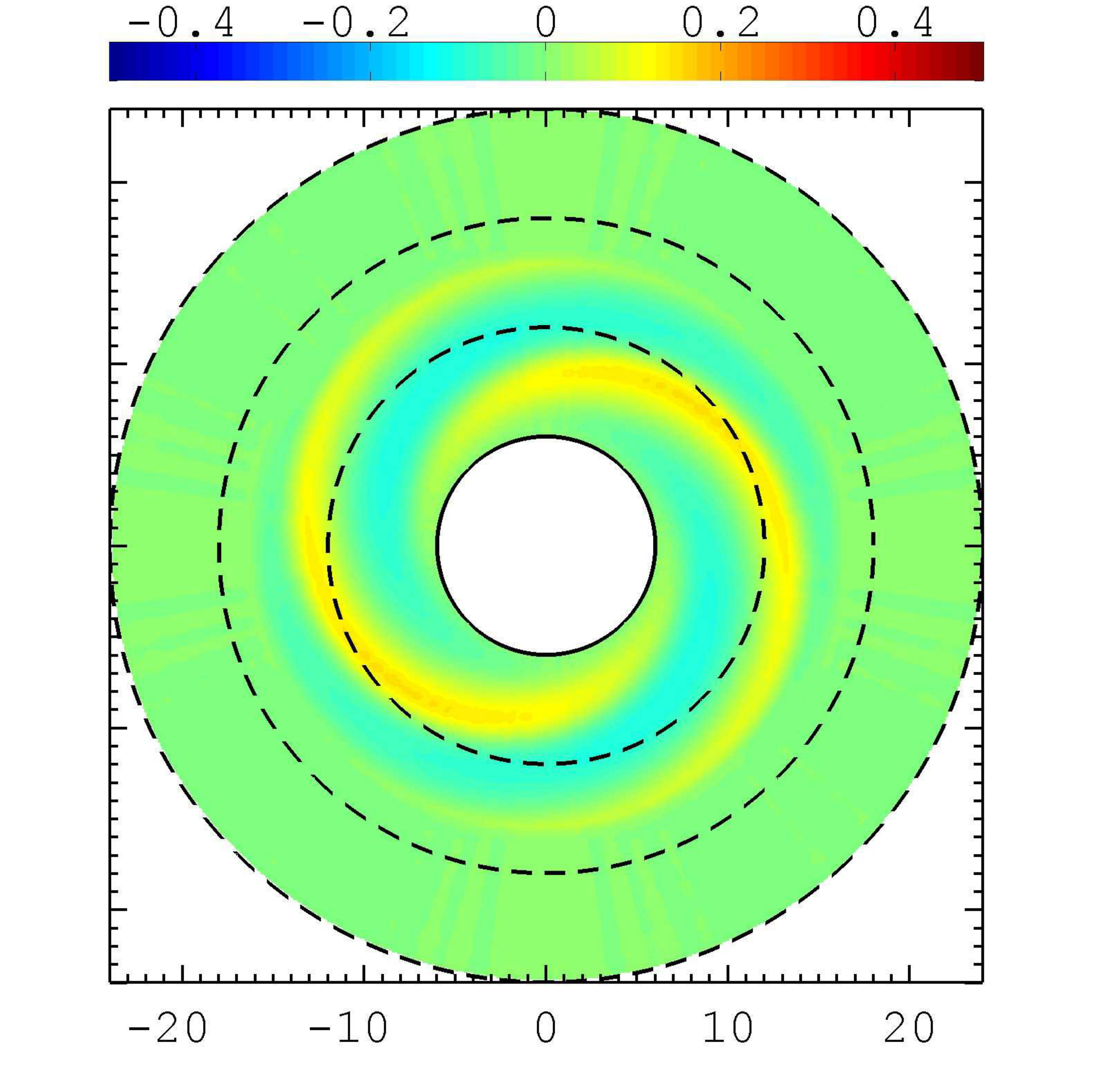}
\includegraphics[width=0.24\hsize]{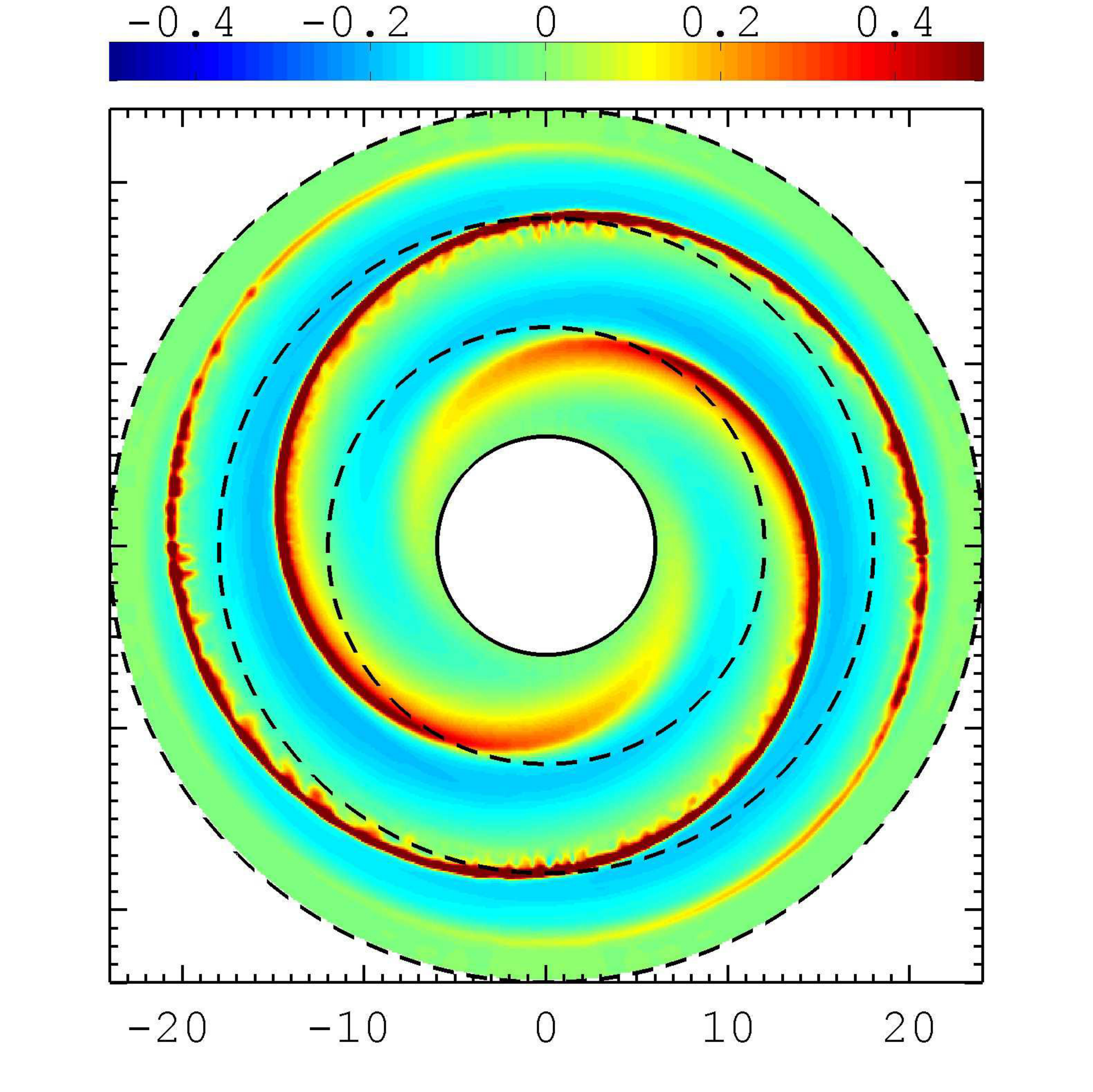}
\includegraphics[width=0.24\hsize]{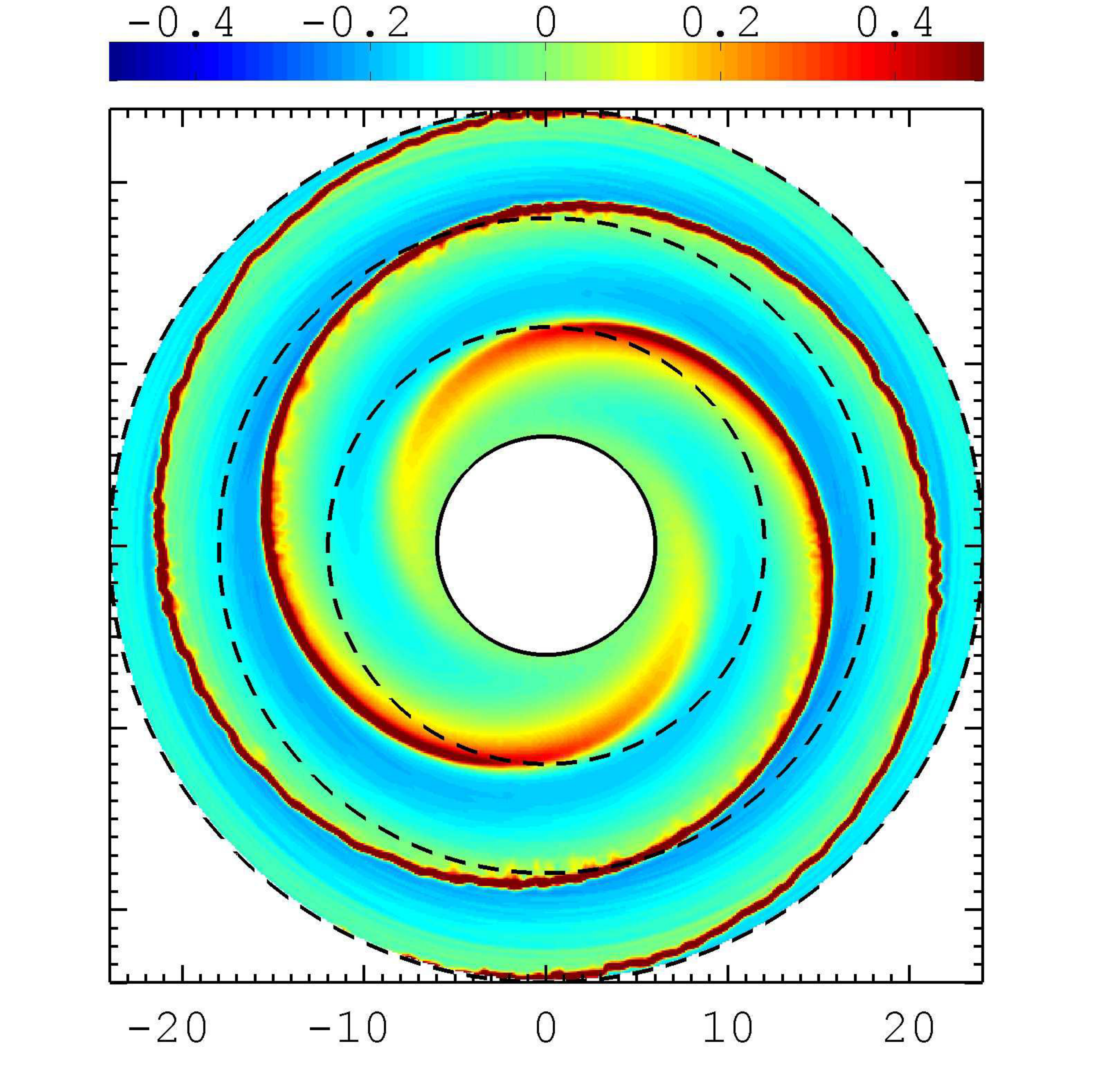}
\includegraphics[width=0.24\hsize]{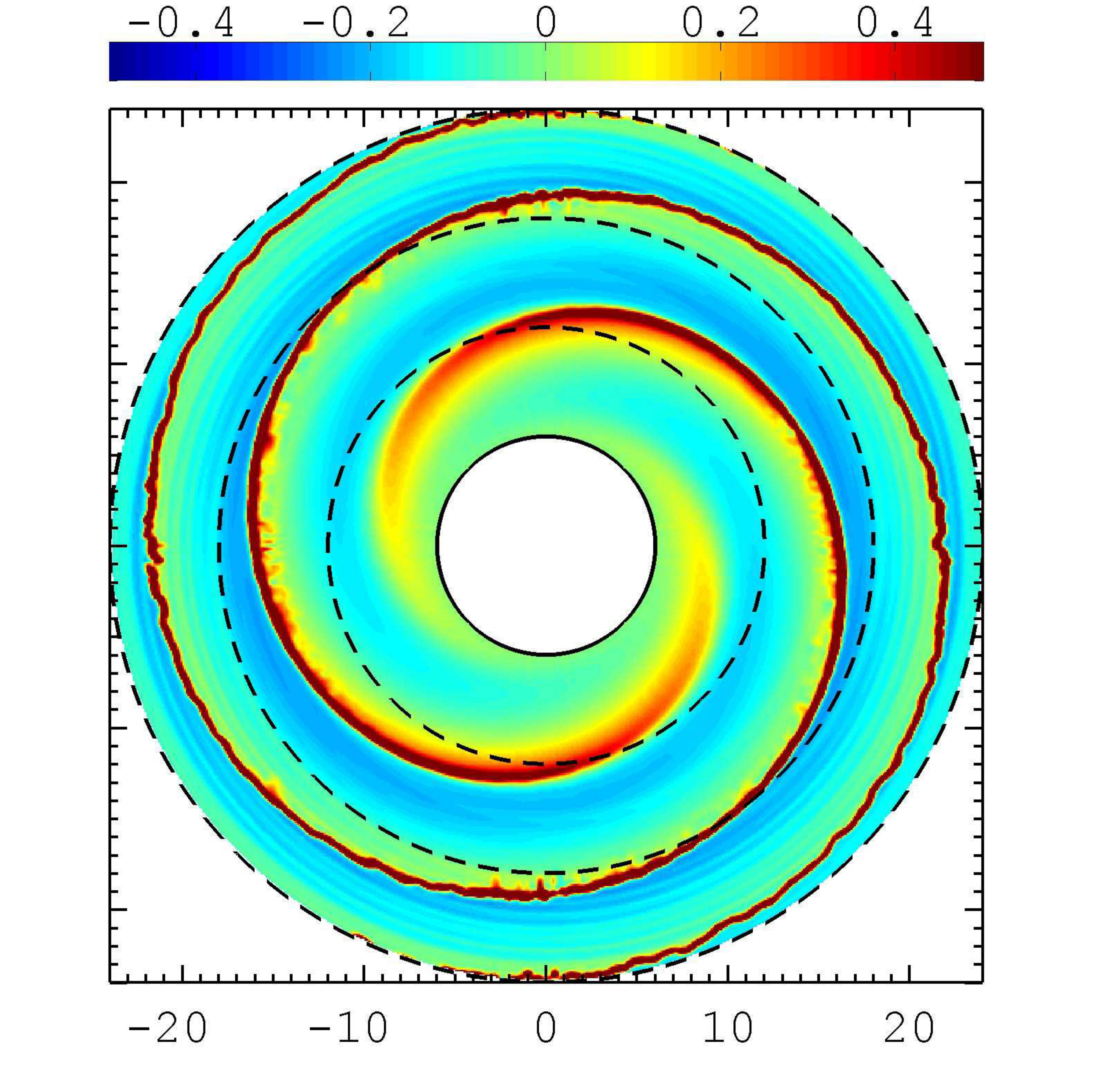}
\includegraphics[width=0.24\hsize]{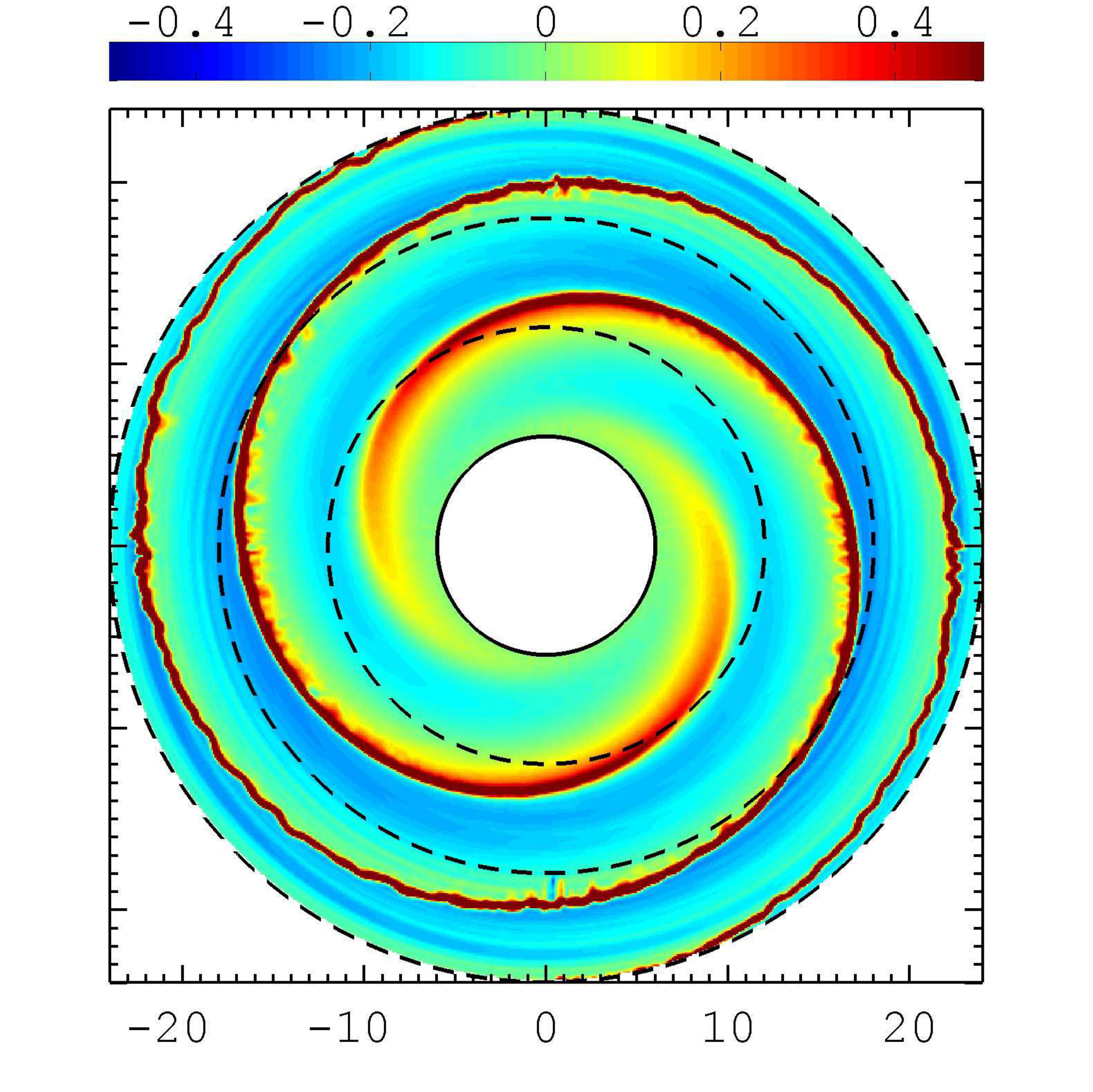}
\includegraphics[width=0.49\hsize]{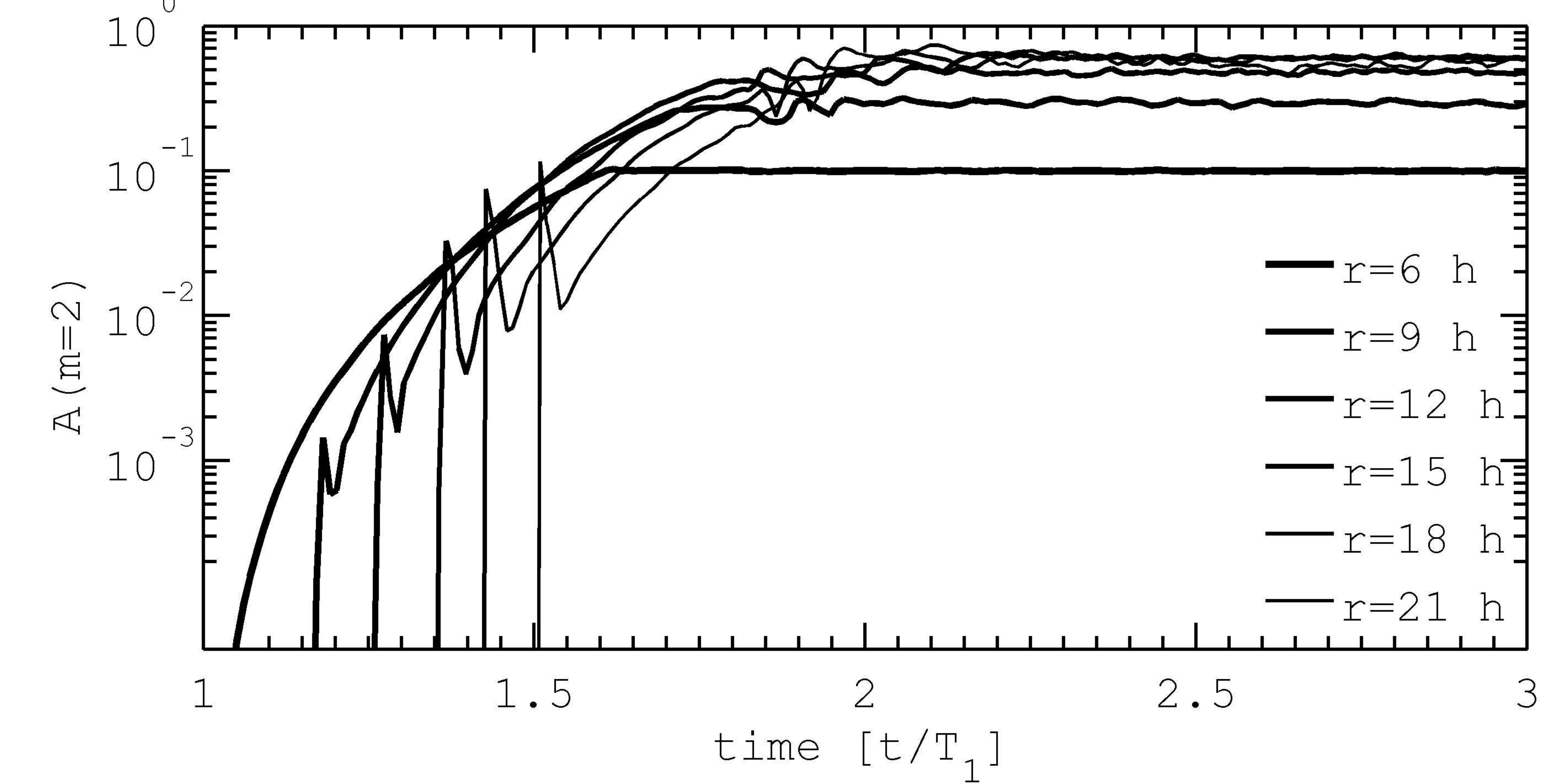}
\caption{Evolution of the relative surface density perturbation ($\Sigma_1/\langle\Sigma_g \rangle$) in model B1 at times $t = 1; 1.1; 1.2; 2; 2.7; 3$ in units of $\approx 500$~Myr.  The plots are drawn in the inertial, nonrotating frame of reference. Black circles are drawn at radii $6h$, $12h$, $18h$, and $24h$. The bottom right panel shows the evolution of the amplitude of the $m=2$ Fourier component of the relative surface density perturbation at different radii, marked by lines with thickness decreasing with increasing radius.}\label{fig::evolution}
\end{figure*}

First we discuss the reference model B1 with azimuthal number $m=2$, pattern speed $\Omega_p = 40$~\kmpskpc, and relative density amplitude $A_0=0.1$. 
In Fig.~\ref{fig::evolution} the evolution of the surface density perturbation is shown. The two-arm trailing spiral structure imposed at the inner boundary moves outwards and its amplitude increases in the outer parts. At time $\approx T_1$ after the beginning of the simulation, a quasi-stationary structure sets in.

\begin{figure}
\includegraphics[width=1\hsize]{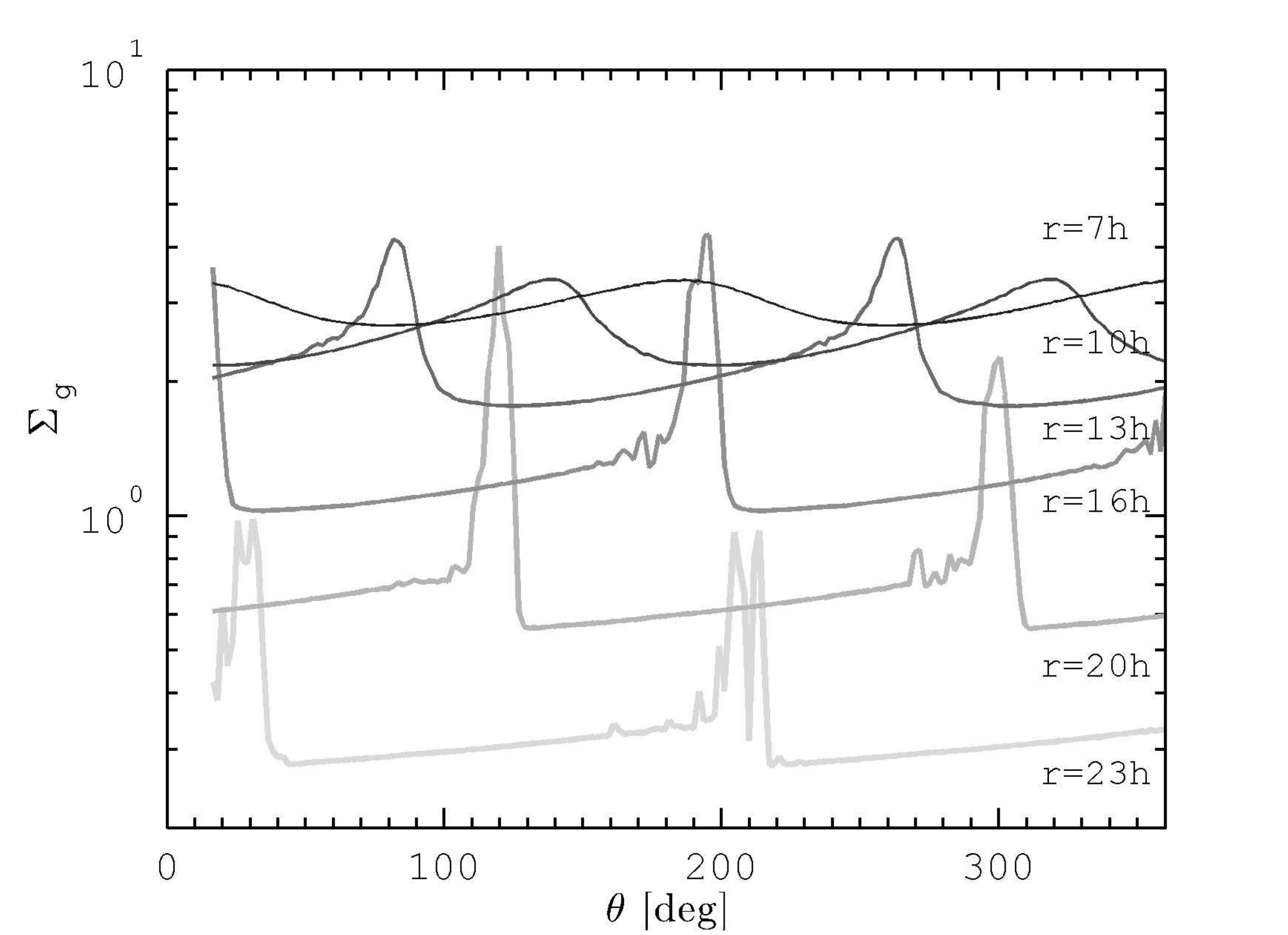}
\caption{Total surface density profiles for the gas ($\Sigma_g$), in units of  \Msun pc$^{-2}$, are shown
along the azimuthal coordinate at given radii for model B1 at $t=2.5T_1$. The line thickness decreases with increasing radius.}\label{fig::shape}
\end{figure}

In the inner regions, inside $\approx 2 r_{\rm opt}$, the spiral density waves are regular and characterized by low relative amplitude $\leqslant 0.1-0.3$. In the outer parts, the wave amplitude increases rapidly; the shape of the density perturbation departs from being sinusoidal and the density perturbation becomes asymmetric~(see Fig.~\ref{fig::shape}). Beyond $\approx 2-2.5 r_{\rm opt}$ rather narrow shocks form. In fact, the Mach number  $\Oo [V(r) - r \Omega_p]/ c$ increases linearly with radius because $c~\approx 3.4$~\kmps and $V(r)\approx210$~\kmps are approximately constant~(see Eq.~(\ref{eq::cr})). Thus the spiral structure is characterized by supersonic motion across the disc. 

In the outermost regions, where the relative amplitude of the spiral density perturbation $\Sigma_1/\langle \Sigma_g\rangle$  reaches the values of $\approx 1-2$, then the shocks become unstable and small-scale spurs appear. It is likely that the instability of spiral shocks makes the perturbation saturate at finite amplitudes. This instability is clearly seen in the evolution of the amplitude of the $m=2$ Fourier component in model B1~(see Fig.~\ref{fig::evolution}). In general, this amplitude remains approximately constant at all radii after the initial transient, that is, at times longer than $t = 1.8-2 T_1$.

The origin of the shock instability is likely to be similar to that of the the wiggle instability investigated numerically by several authors~\cite[e.g.,][]{2004MNRAS.349..270W,2006MNRAS.367..873D,2014ApJ...789...68K}. In other words, Kelvin-Helmholtz instability is likely to be one important mechanism for the formation of spurs in the vicinity of the spiral shocks. The numerical simulations by~\cite{2004MNRAS.349..270W} suggest that tightly-wound spiral shocks should be relatively stable, compared with the case of open spirals. However, in our simulations the large amplitude of the shock associated with the observed instability occurs in the outermost parts of the disc ($r > 12h\approx2 r_{\rm opt}$), where the pitch angle is smaller than $5^{\circ}$~(the radial profile of the pitch angle $i(r)$ is illustrated in Fig.~\ref{fig::lincomp}). In turn, in the inner parts of the computational domain ($r<12h$) the pitch angle is close to $10^\circ$; but there the amplitude of the gas perturbation is relatively small so that the density waves are in the linear regime~(see also the caption to Fig.~\ref{fig::shape}) and the strong shock instability is suppressed. It should be noted that in our simulations the spatial size of spurs is about one kiloparsec, which is comparable to the size of some small-scale structures inside the optical radius of nearby galaxies. 

 \begin{figure}
\includegraphics[width=1\hsize]{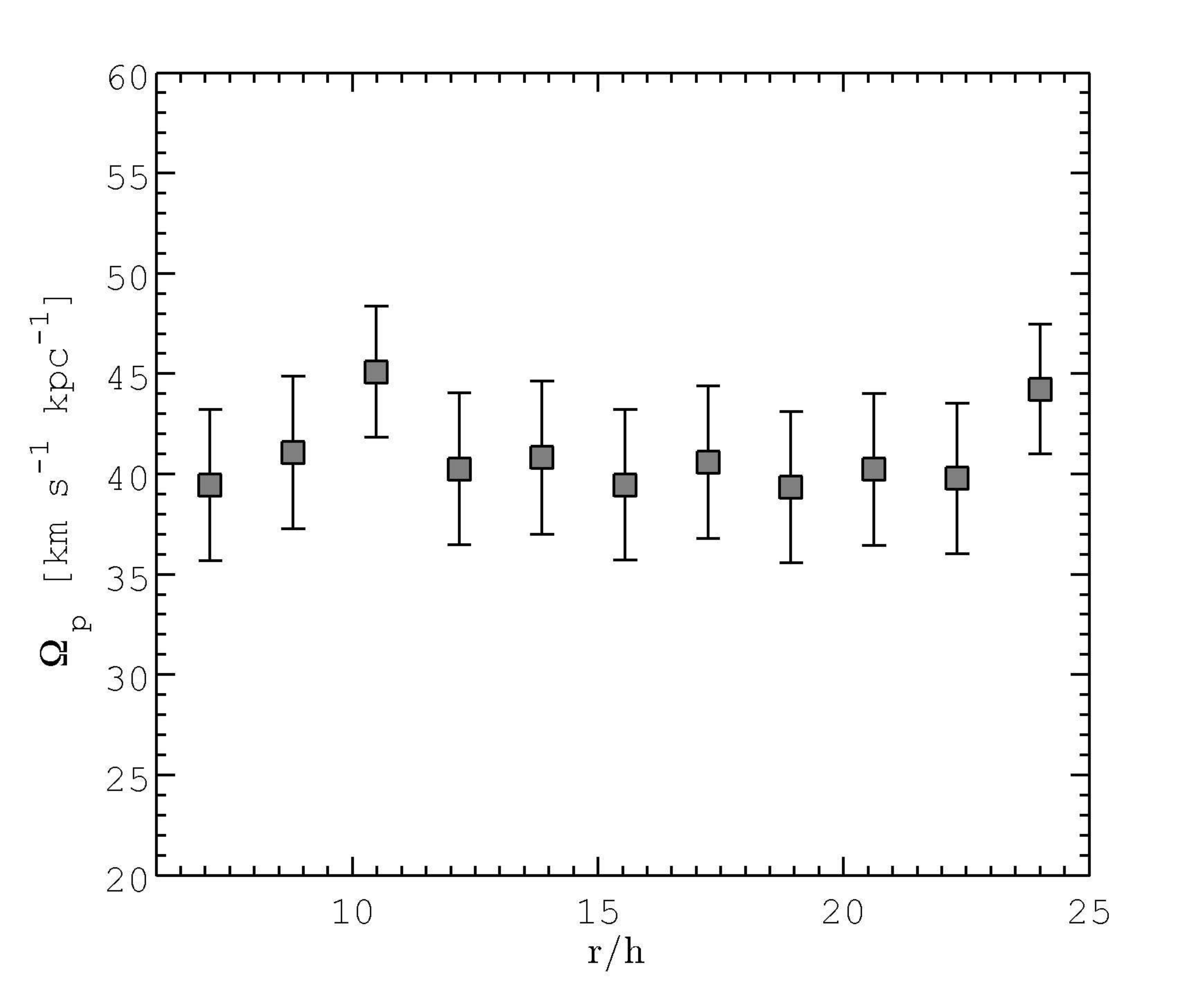}
\caption{Directly measured pattern speed at different radii for model B1.}\label{fig::pattern_speed}
\end{figure}

\subsection{Direct measurements of pitch angles and pattern speeds}\label{sec::measurements}
To derive the radial profile of the pitch angle of the spiral arms observed in our simulations, we have used the method described by~\cite{2012ApJS..199...33D}. We consider the Fourier analysis of the surface density distribution of the perturbation $\Sigma_1$ as a function of azimuth $\theta$ and the Fourier transform in the logarithmic radial coordinate $u=\log{r}$ (this is often referred to as a decomposition in logarithmic spirals). The integrations are performed over narrow annular rings in the computational domain, that is:
\begin{equation}\label{eq::pitch_four_amp}
 \Oo A(p, m, r_i) = \frac{1}{G_0} \int_{u_{r_{i}}}^{u_{r_{i+1}}} \int_{-\pi}^\pi \Sigma_1(u, \theta) {e}^{- {i}(m\theta + pu)} {d}\theta  {d}u\,,
\end{equation}
where   $\Oo G_0 =  \int_{u_{6h}}^{u_{24h}} \int_{-\pi}^\pi \Sigma_1(u, \theta) {d}\theta  {d}u$ is a suitable normalization constant. Thus $A(p, m , r_i)$ represents the contribution of the $m$-armed logarithmic spiral component, with pitch angle $i=\arctan\left(-m/p\right)$ at given radius $r_i$. By considering the value $p_{max}$ at which the quantity $A(p, m , r_i)$ attains its maximum at given $m$ and $r_i$ we can thus reconstruct the pitch angle radial profile for our simulated spiral structures. The error on the pitch angle value $i_i$ can be found from the spatial variation of the quantity in the range $[r_{i-1/2}; r_{i+1/2}]$.

In the narrow annuli at radius $r_i$ considered in the method just described, we can also measure the pattern speed associated with the spiral structure present. To make such measurement, we proceed as follows.  A phase angle for given $m$ and $p$ defining the orientation of the spiral pattern at radius $r_i$ can be calculated as 
\begin{equation}
\Psi = \arctan \frac{ {\rm Im} (A)}{{\rm Re} (A)}\,,
\end{equation}
where ${\rm Im}(A)$ and ${\rm Re}(A)$ are the imaginary and the real part of $A(p_{max}, m)$, respectively. Then a local value of the speed of the pattern with given $m$ can be determined as 
\begin{equation}\label{eq::omega_p}
\Oo \Omega_p = \frac{1}{m}\frac{\partial \Psi}{\partial t}\,,
\end{equation}
The error on the value of the pattern speed depends on both the spatial variation of the derivative in~Eq.~(\ref{eq::omega_p}) and on small time-dependent variation of the quantity that is calculated. When a single mode is imposed at the inner boundary, the procedure indeed gives back the value of the pattern speed of the imposed perturbation.  In Fig.~\ref{fig::pattern_speed}, calculated at $t \approx 2 T_1$ for the B1 model, the measured pattern speed is shown to be constant with radius and consistent with that of the single-mode perturbation imposed at the inner boundary.

\subsection{Comparison with the linear theory}\label{sec::linear}
From the linear theory of density waves~\cite{2010A&A...512A..17B} obtained the expressions for radial velocity, surface density, and pitch angle of short-trailing density waves in the galactic outer regions~\cite[e.g., see Eqs.~(9) and (13)~in][]{2010A&A...512A..17B}. In this section we compare the results of our dynamical simulations with the predictions of the linear analysis.

 \begin{figure}
\includegraphics[width=1\hsize]{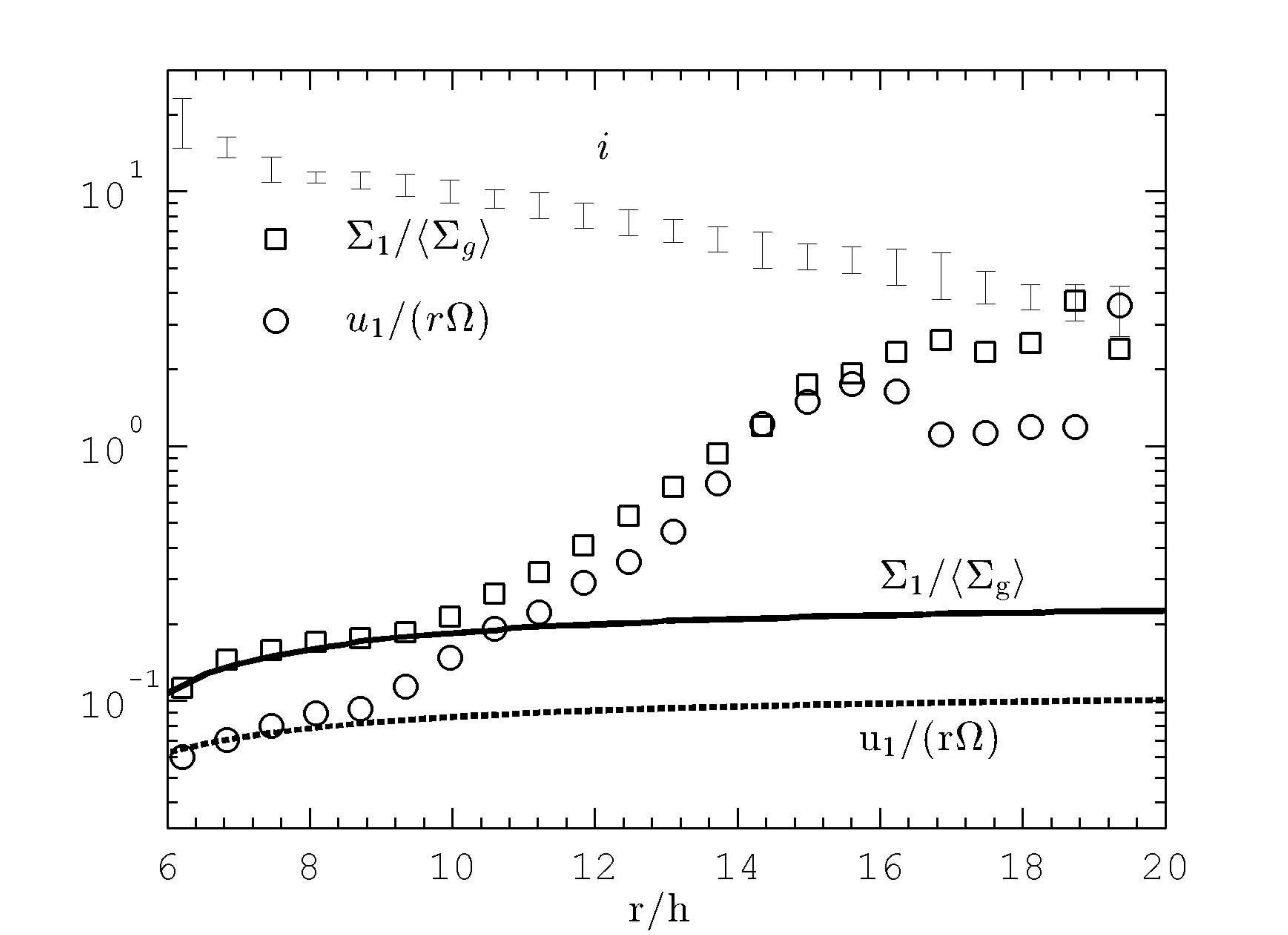}
\caption{The relative amplitude of the density wave in model B1 is shown by squares. Circles represent the radial velocity perturbation relative to the rotation velocity of the basic state for the same model. The radial profile of the pitch angle in model B1 is shown by error bars; error bars are associated with the averaging of the pitch angle in rings of finite radial size~(see~Eq.~\ref{eq::pitch_four_amp}). The solution from the linear theory by~\citet{2010A&A...512A..17B} is shown by lines. }\label{fig::lincomp}
\end{figure}

For the frame at $t = 3 T_1$ from the reference B1 simulation~(see bottom row in Fig.~\ref{fig::evolution}) we calculate the radial profile of surface density perturbation,  radial velocity perturbation, and pitch angle~(see~Fig.~\ref{fig::lincomp}). The perturbation amplitudes of surface density and radial velocity increase with radius, whereas the pitch angle of the pattern decreases. This qualitative behaviour agrees well with the expectations of the linear theory~\cite[see~Fig.~5 in][]{2010A&A...512A..17B}. Quantitatively,  the good agreement between linear theory and simulations applies to a rather wide radial range $6h<r<10-12h$. At larger radii ($r>2 r_{opt}$), the simulations exhibit a strongly nonlinear behaviour, because the relative perturbation amplitudes attain very high values, up to $2-3$. 
 
\subsection{Model with higher velocity dispersion}\label{sec::high_c}

In this section we consider a different model (B7) characterized by a velocity dispersion higher than that adopted in the reference B1 model, well above the value required by the condition of marginal axisymmetric stability. We thus take $c = 5$ km~s$^{-1}$. Interestingly, the dynamical evolution
of this hotter system basically follows the same picture as described in
Sect.~\ref{sec::single_mode}. Of course some morphological changes are expected and indeed found in the simulations.
 
In Fig.~\ref{fig::m2diff} we show the relative surface density distribution established at time $2.5~T_{1}$. Because of the higher gas velocity dis-
persion, in the B7 model we observe the excitation of a significantly
more open spiral structure (with respect to the reference B1 model).
In the regions close to the inner boundary, the pitch angle is $\approx 20^\circ$ (to be compared to the value of $\approx 10^\circ$ found in B1). Because a larger pitch angle of the pattern provides better conditions for the shear instability of the spiral shocks, a more perturbed morphology of the pattern and the presence of prominent spurs and feathers along the spiral arms are expected and found in the simulations.

\subsection{More realistic models}\label{sec::phys}
Our reference model~B1 demonstrates the possibility of regular and sharp spiral patterns, of the type that is observed in some deep HI images~(e.g.,~in NGC~1512), but physically it is exceedingly simple. More realistic models should be devised.  In particular, we have checked how different small-scale processes, which can be implemented at the subgrid level in our simulations, affect the morphology of the observed spiral structure. 

\begin{figure*}
\includegraphics[width=0.33\hsize]{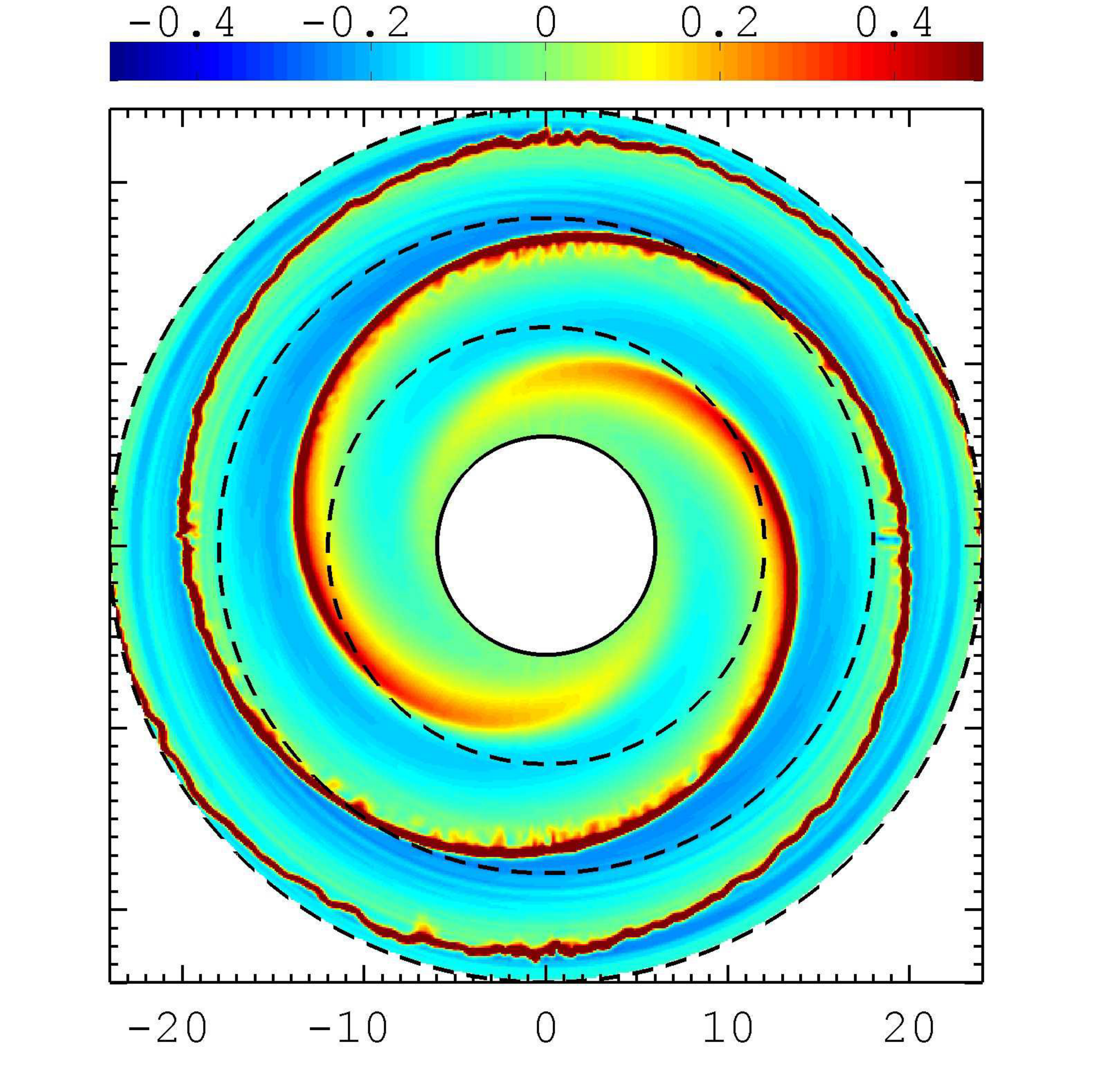}
\includegraphics[width=0.33\hsize]{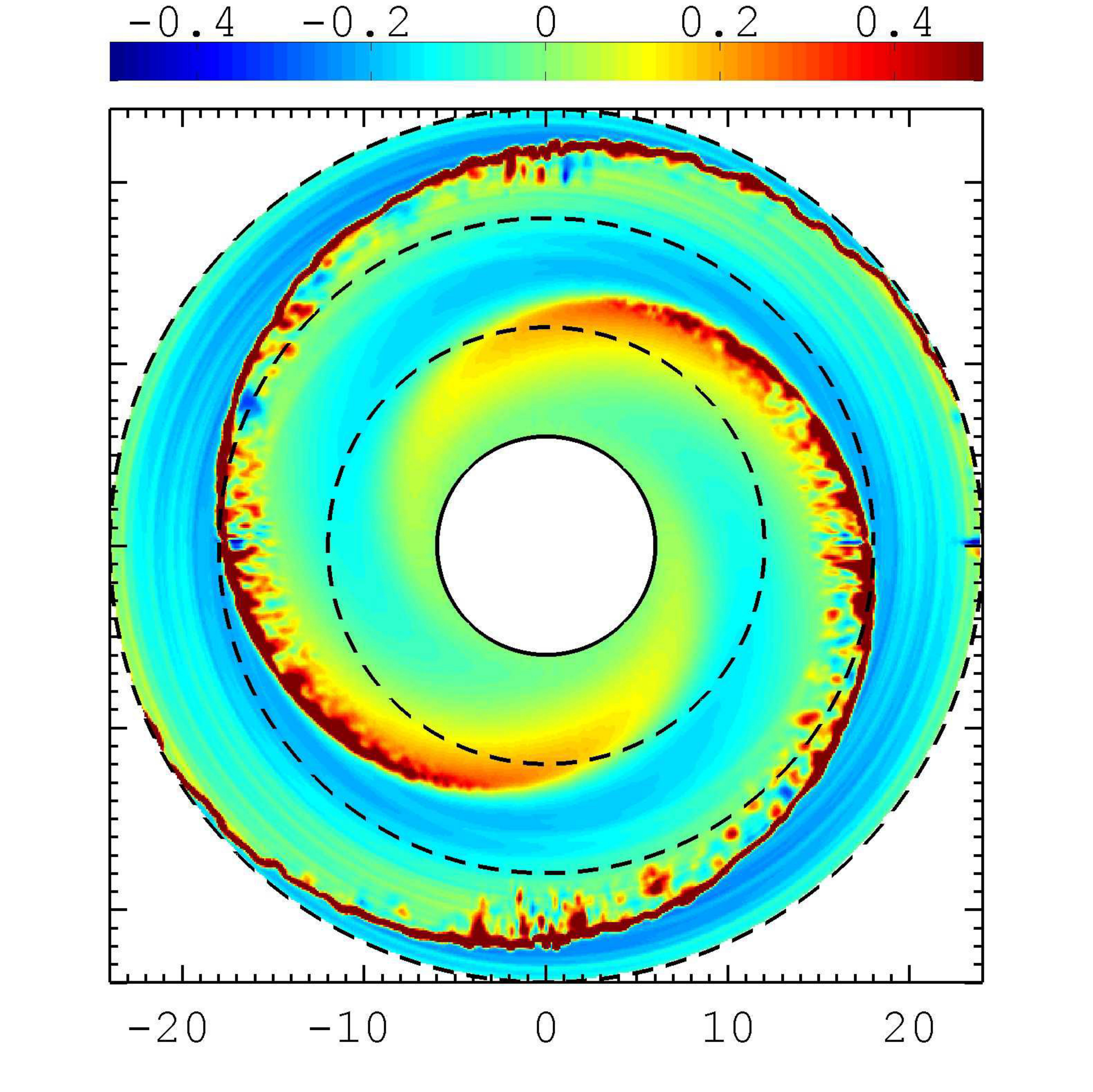}\\
\includegraphics[width=0.33\hsize]{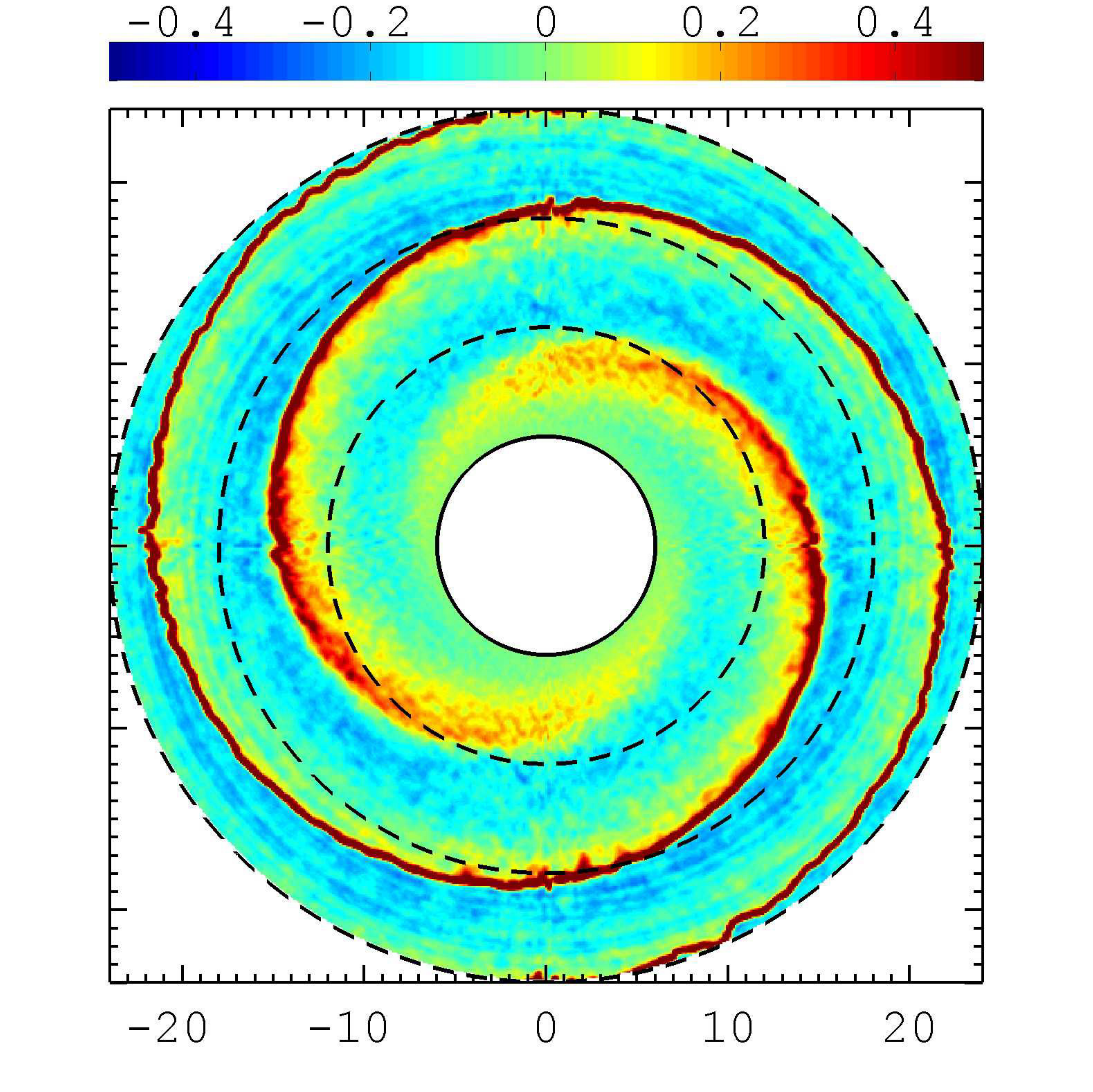}
\includegraphics[width=0.33\hsize]{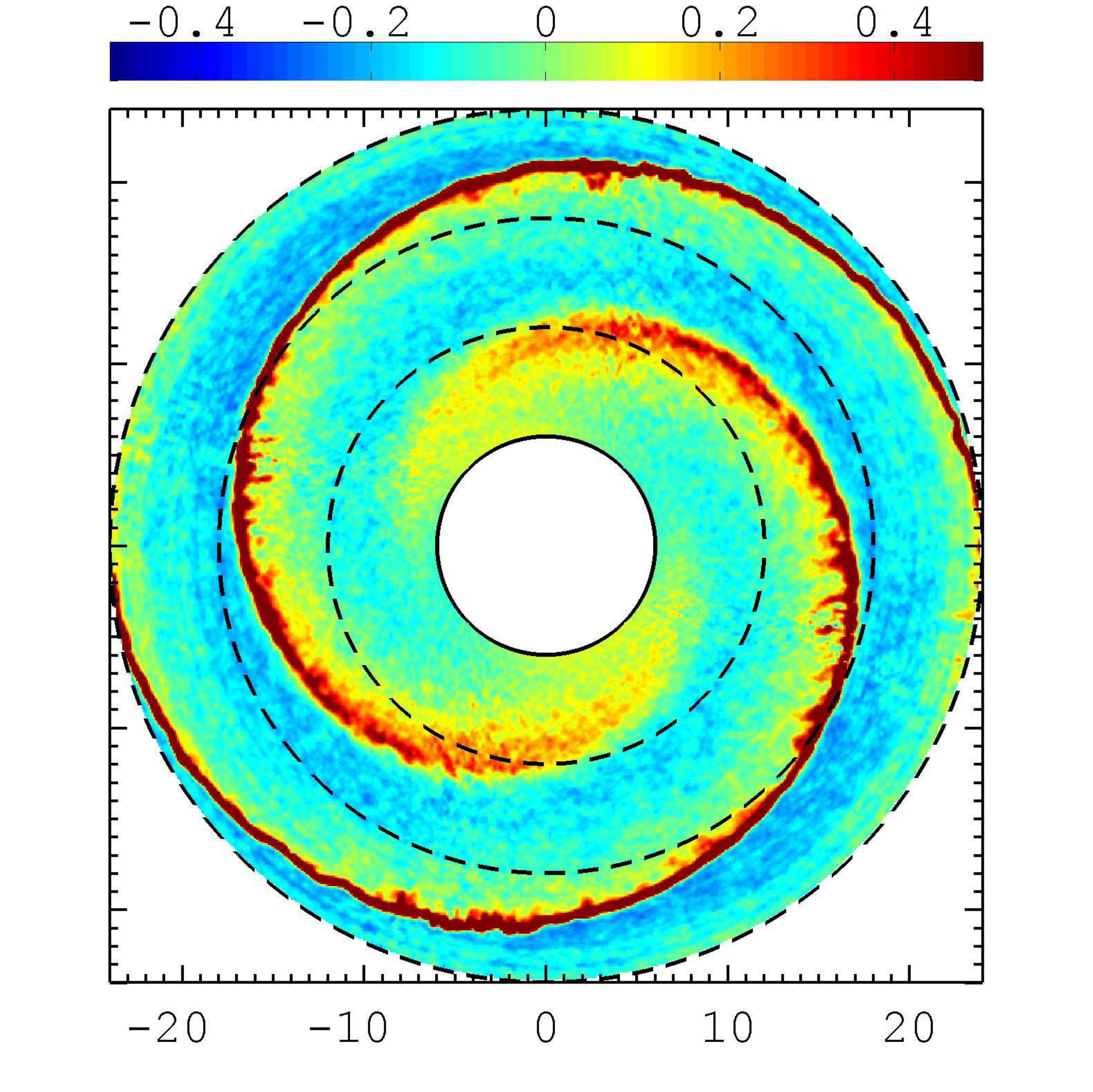}
\includegraphics[width=0.33\hsize]{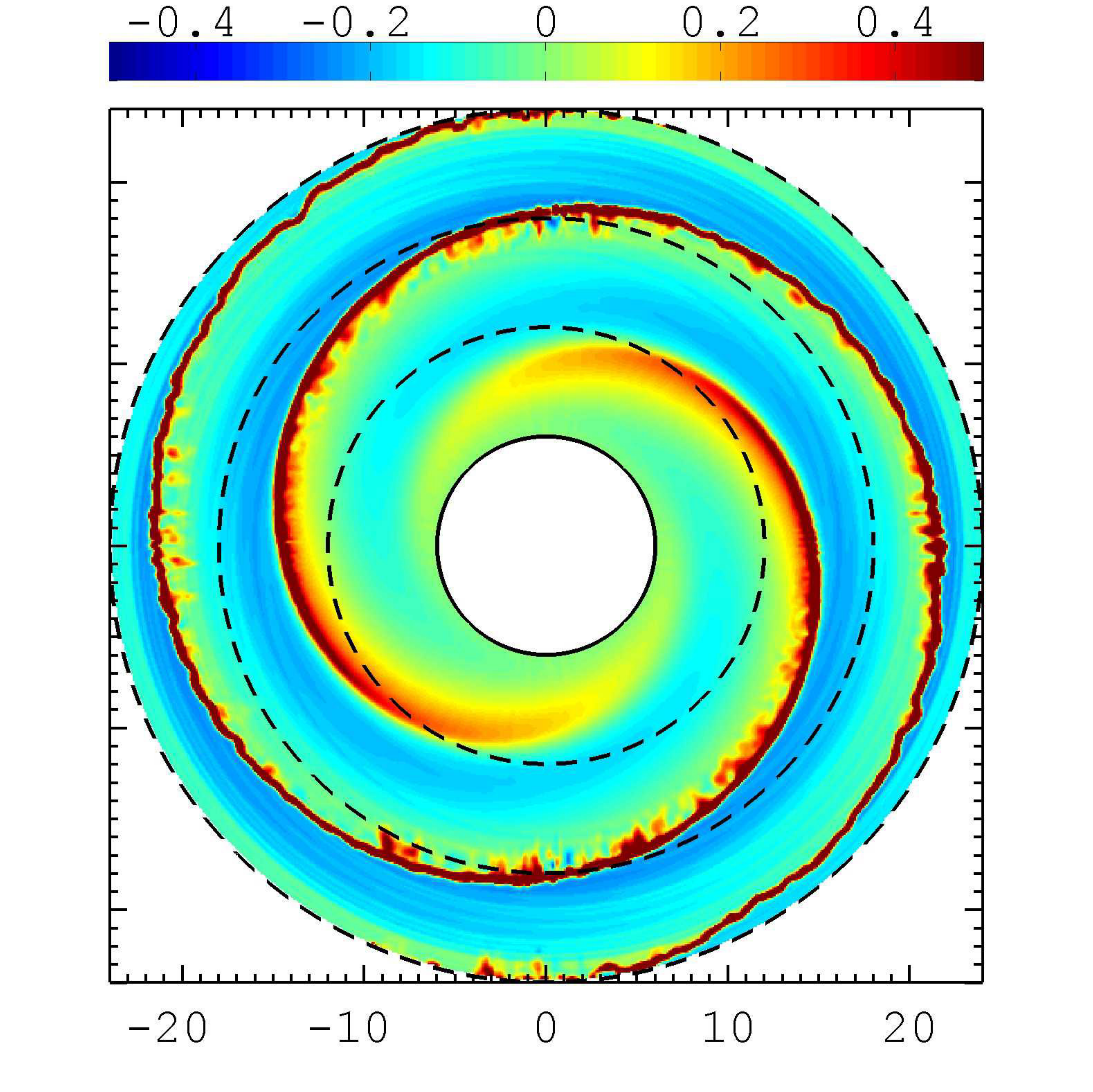}
\caption{Relative surface density perturbation ($\Sigma_1/\langle\Sigma_g \rangle$) for different models, with a single $m=2$ perturbation imposed at the inner boundary, at time $t = 2.5T_1$. From left to right: top row, B1~(the reference model) and B7 (with higher gas velocity dispersion); bottom row, B2~(with a clumpy initial gas distribution), B4~(with a clumpy dark halo), B5~(with subgrid cooling). For a detailed parameter description, see Table~\ref{tab::ini}. Black  circles mark different radii, at $6h$, $12h$, $18h$, and $24h$.}\label{fig::m2diff}
\end{figure*}

Here we compare four types of models with the same imposed perturbation at the inner boundary: (i) the reference B1 model, which was described previously; (ii) the B2 model, which is based on a clumpy gas distribution and a smooth potential; (iii) the B4 model, which is based on a smooth gas distribution and a halo potential perturbed by clumps of dark matter; and (iv) the B5 model, which is the same as B1, but takes subgrid cooling into account. Results of simulations for these models are shown in Fig.~\ref{fig::m2diff}, where the gas surface density perturbation $\Sigma_1$ is illustrated at $t \approx 2.5T_1$. The basic conclusion is that the results shown for the B1 model are robust. 

In fact, the large-scale morphology (grand design) is similar for all models. However, various additional features are found to characterize the small-scale morphology. For the B2 and B4 models the new small-scale features basically cover the entire disc. In contrast, in the B5 model, with subgrid cooling, the effects are most evident in the denser regions at the edge of the spiral shocks, making the arms less smooth; the intensity and spatial scales of the spurs and feathers that are observed in the simulations suggest that the B5 model involves the wiggle instability known to affect a multi-phase inhomogeneous interstellar medium in the presence of spiral shocks on the galactic scale~\citep{2008ApJ...675..188W}.   

\subsection{Simulations with more than one mode imposed at the inner boundary}\label{sec::several_modes}
\begin{figure*}
\includegraphics[width=0.325\hsize]{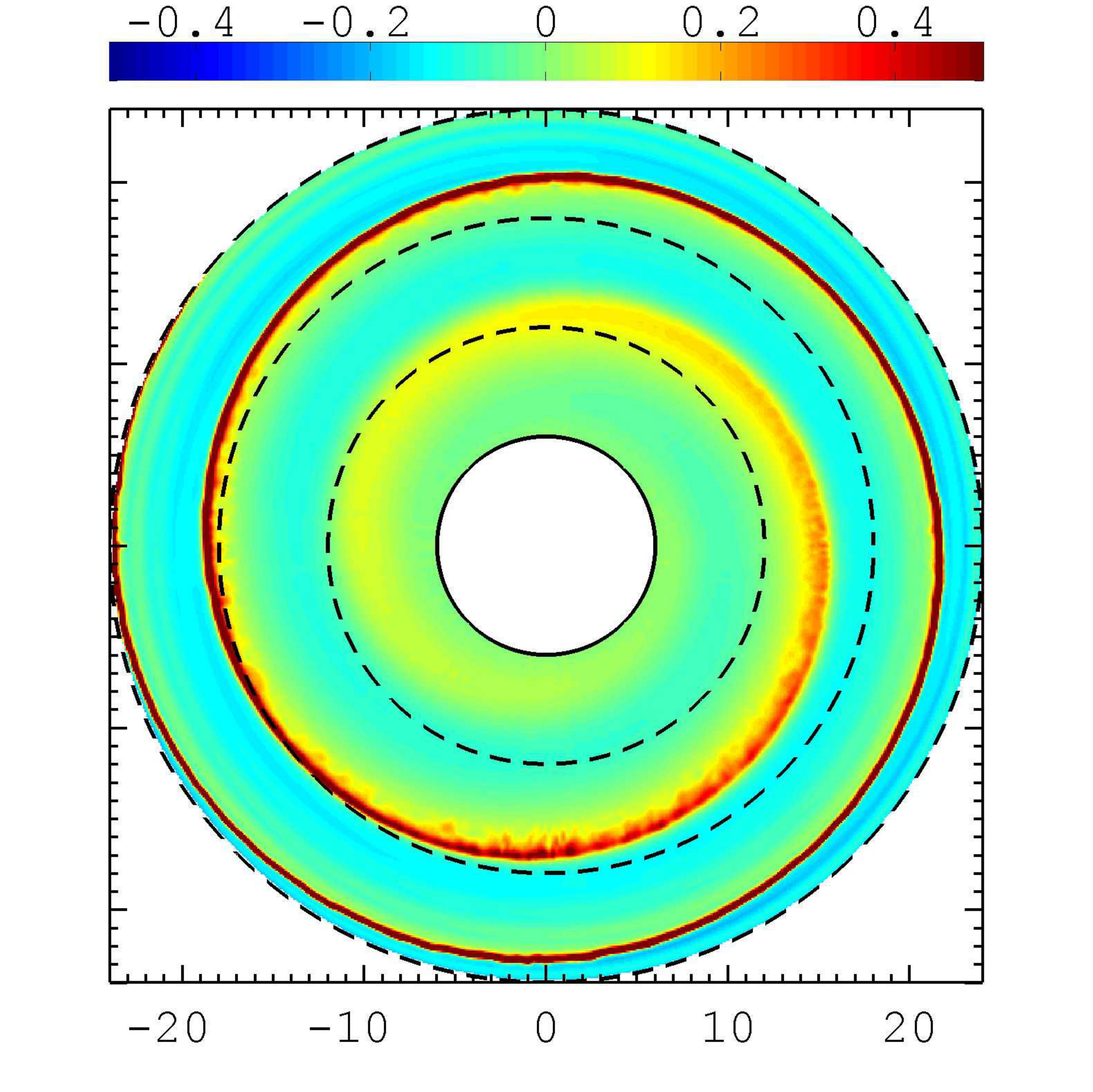}
\includegraphics[width=0.325\hsize]{B1_Sigma2D0018b.eps}
\includegraphics[width=0.325\hsize]{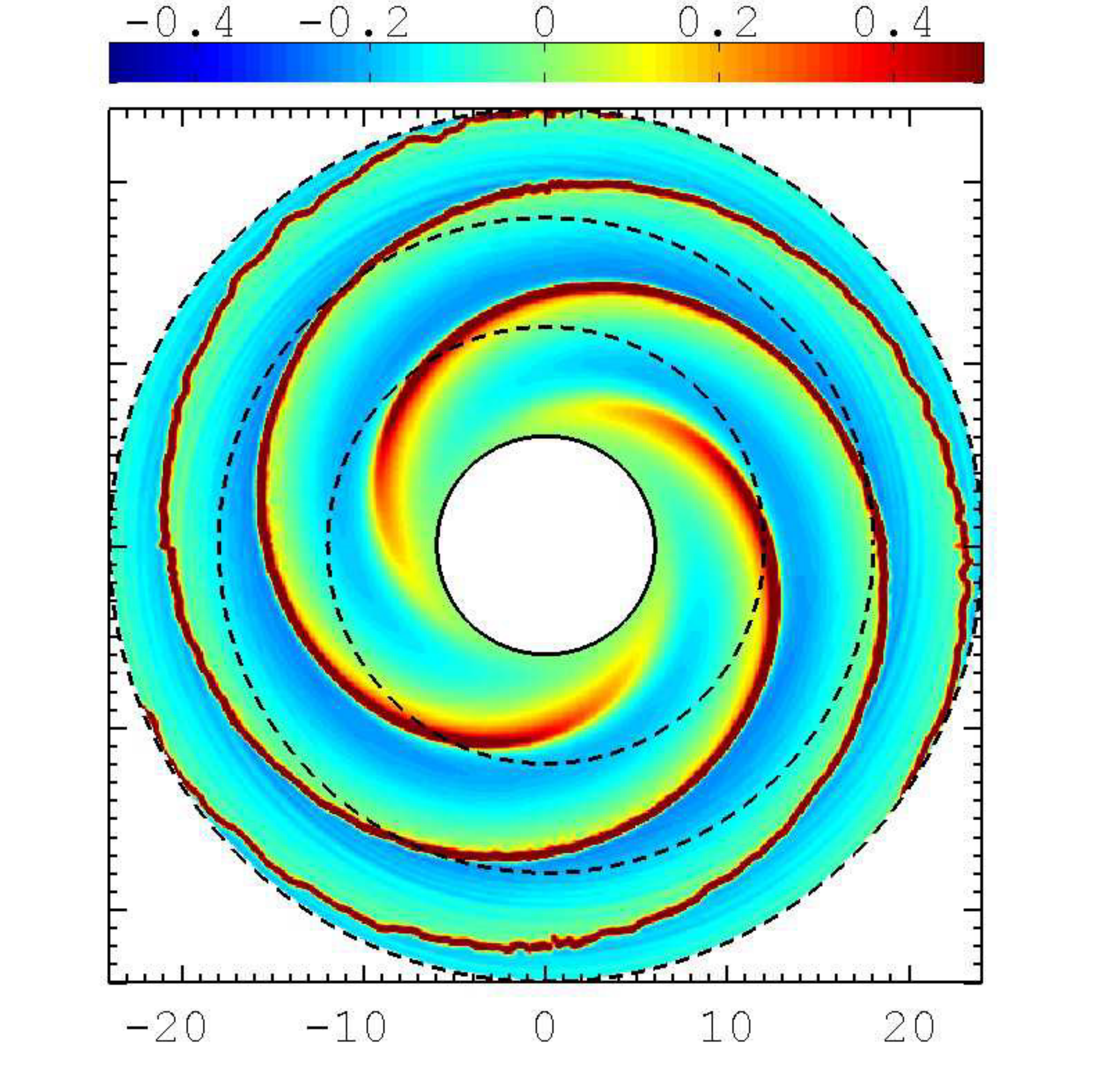}
\includegraphics[width=0.325\hsize]{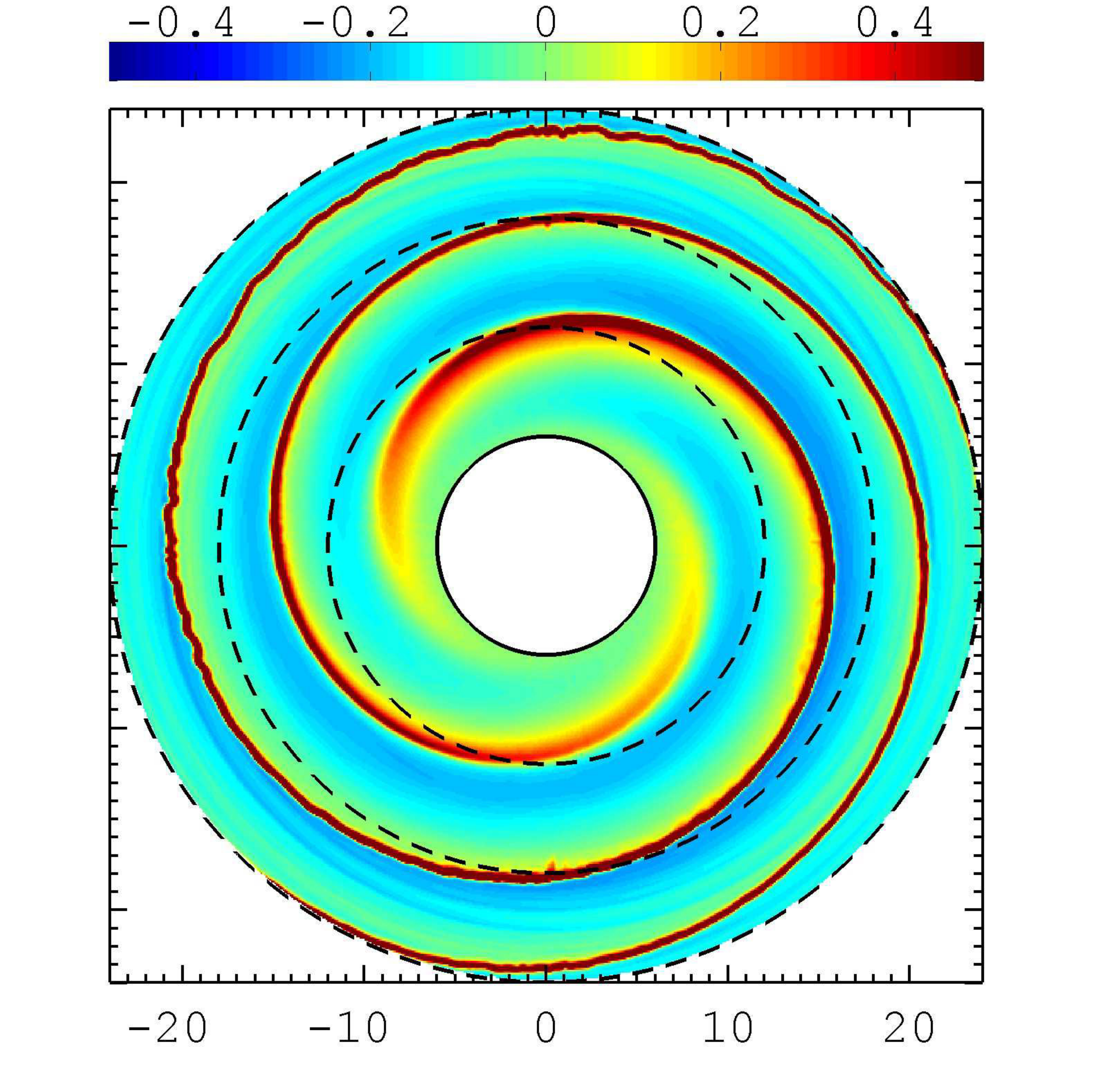}
\includegraphics[width=0.325\hsize]{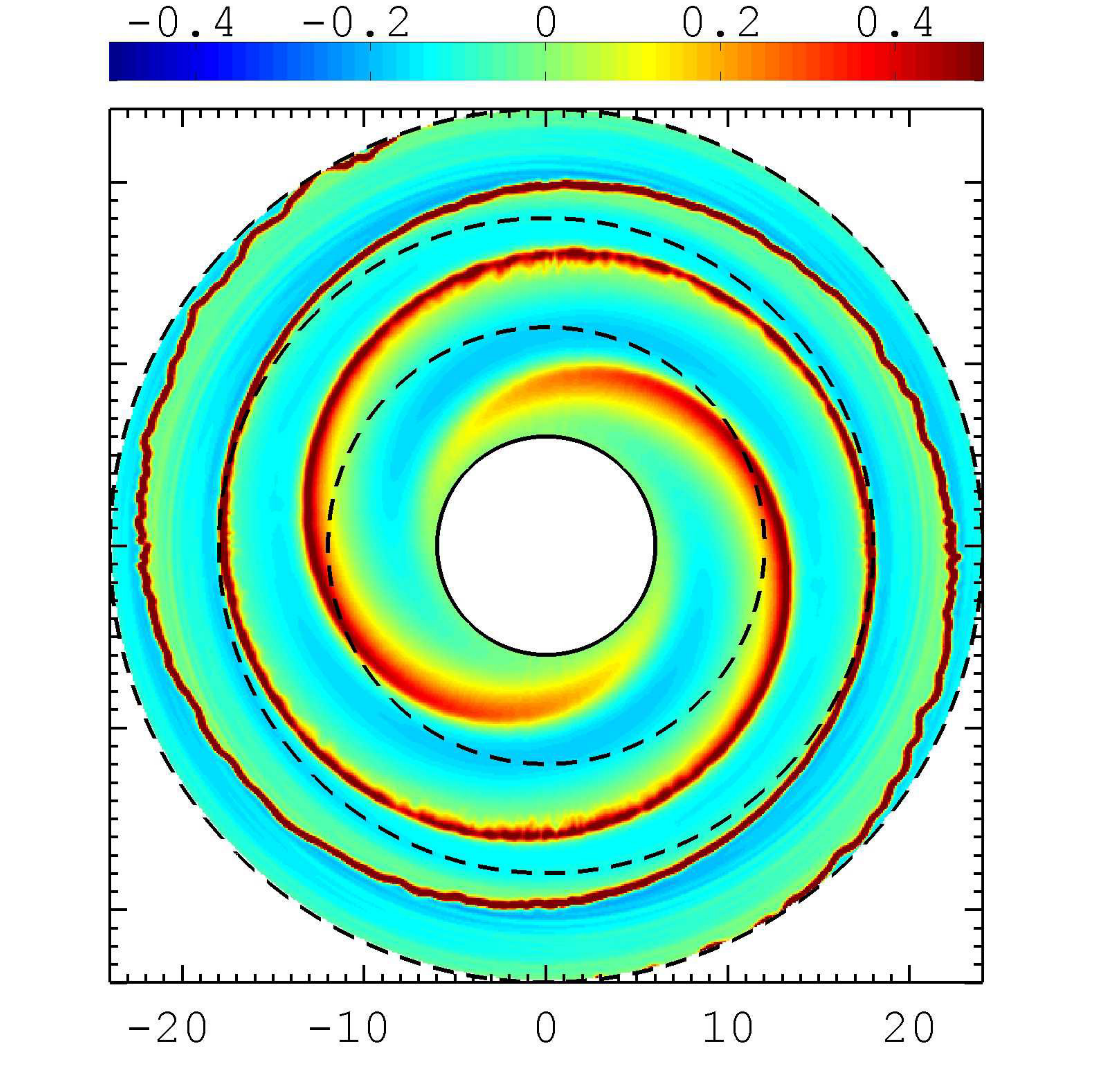}
\includegraphics[width=0.325\hsize]{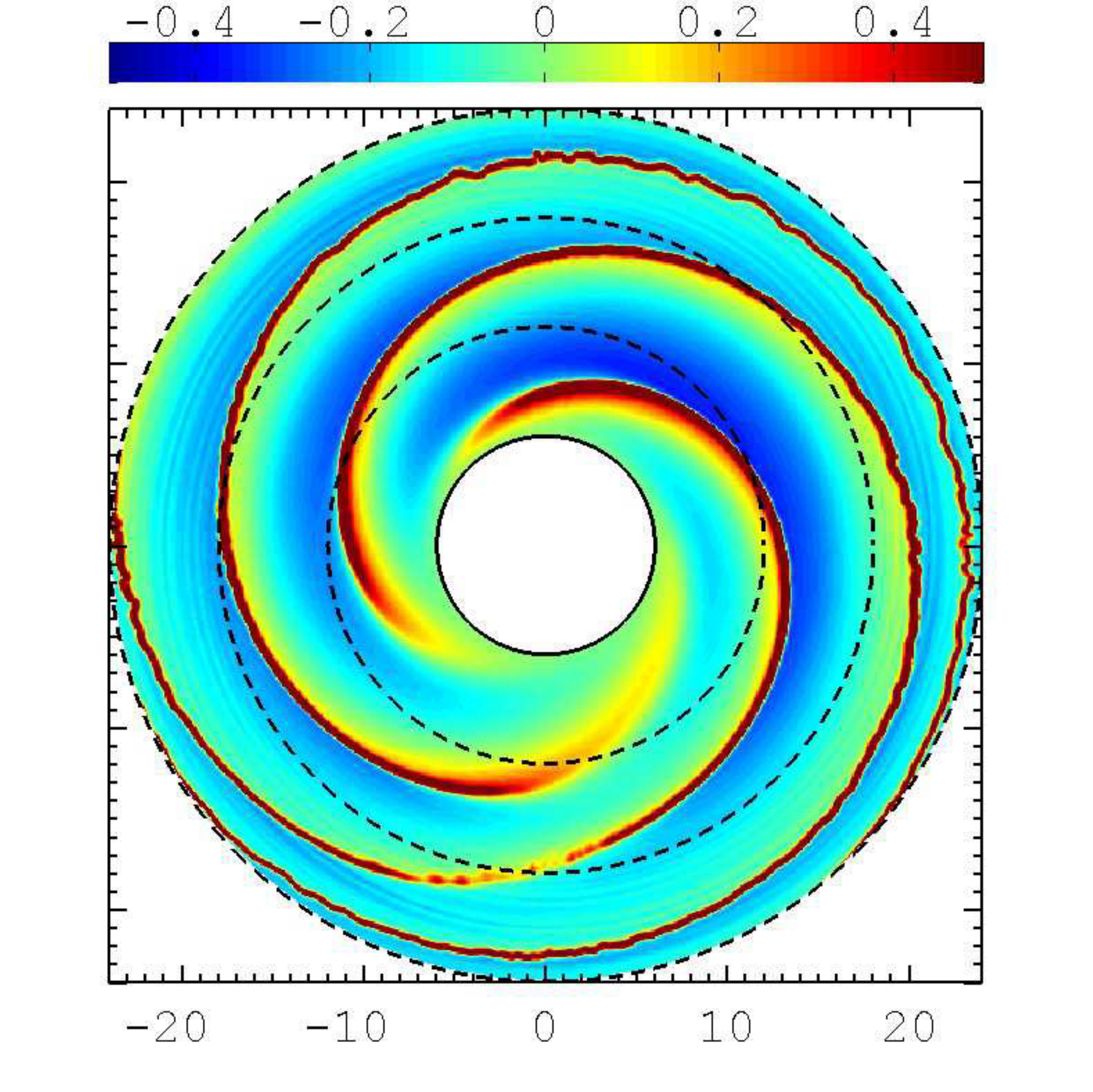}
\caption{Relative surface density perturbation ($\Sigma_1/\langle\Sigma_g \rangle$) for models with different conditions imposed at the inner boundary, at time $t = 2.5T_1$. Top row: E1~($m=1$), B1~($m=2$), K1~($m=3$). Bottom row: H1~(superposition of an $m=1$ and an $m=2$ mode), F1~(superposition of a pair of $m=2$ modes), J1~(superposition of an $m=1$, an $m=2$, and an $m=3$ mode) (see Table~\ref{tab::ini}). Black circles mark different radii, at $6h$, $12h$, $18h$, and $24h$.}\label{fig::suppos}
\end{figure*}

So far we have described models on which a single mode is imposed at the inner boundary. In real grand-design spiral galaxies, it is natural to expect that the large-scale morphology is dominated by the superposition of few spiral modes, each characterized by its own amplitude and pattern speed. We thus investigate the properties of simulations in which at the inner boundary a superposition of several non-axisymmetric modes is imposed. The H1 model is based on the combination of an $m=1$ mode rotating at $30$~\kmpskpc and an $m=2$ mode rotating at $40$~\kmpskpc. The F1 model considers the superposition of a pair of $m=2$ modes with different amplitudes ($A_{0,1} = 0.05$, $A_{0,2} = 0.1$) and pattern speeds ($30$~\kmpskpc, $40$~\kmpskpc). The J1 model studies the case in which three modes with different $m$ numbers are present ($m=1$, $m=2$, and $m=3$). A description of the adopted parameters is given in Table~\ref{tab::ini}. Figure~\ref{fig::suppos} illustrates the surface density perturbation maps for the single-mode cases, that is, the standard B1 model ($m=2$), the E1 model ($m=1$), and the K1 model~($m=3$), and for the models with several modes (H1, F1, J1).

The evolution of the disc in the case in which several modes are imposed at the inner boundary is rather similar to that of the single-mode case. The spiral morphology of discs where a dominant $m = 3$ perturbation is applied tend to exhibit a prominent three-armed structure during the simulation.  The presence of an $m=1$ imposed perturbation is generally associated with some lopsidedness. The models are characterized by a time-dependent evolution of spiral structure, although the overall observed patterns do not appear to vary very significantly in time. Of course, the evolution is associated with the superposition of the modes, rotating with different angular speeds. The qualitative behaviour remains generally similar to that of the B1 model, with nonlinear behaviour setting in at large radii and some wiggle instability occurring when subgrid cooling is incorporated.

\begin{figure*}
\includegraphics[width=0.24\hsize]{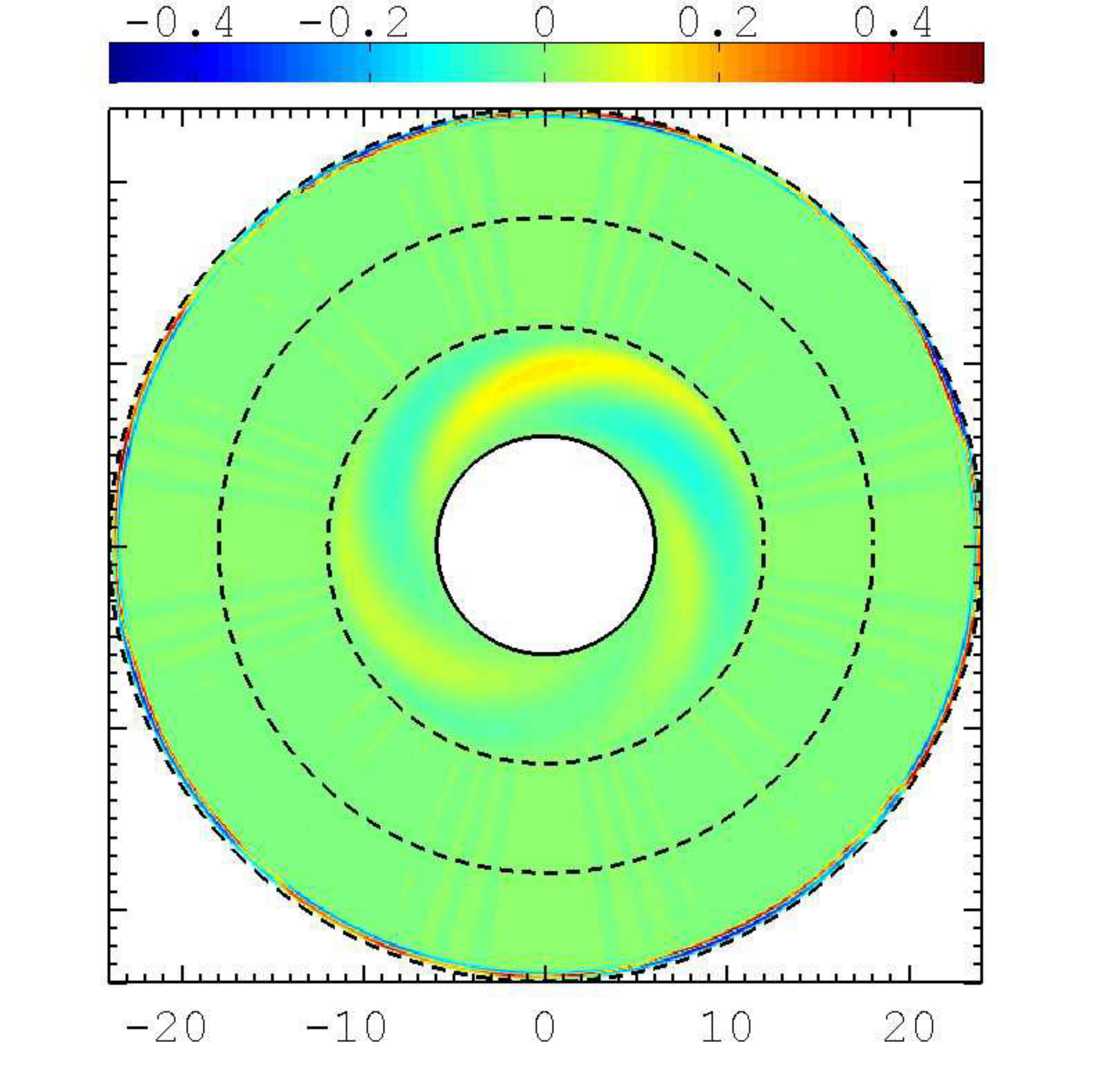}
\includegraphics[width=0.24\hsize]{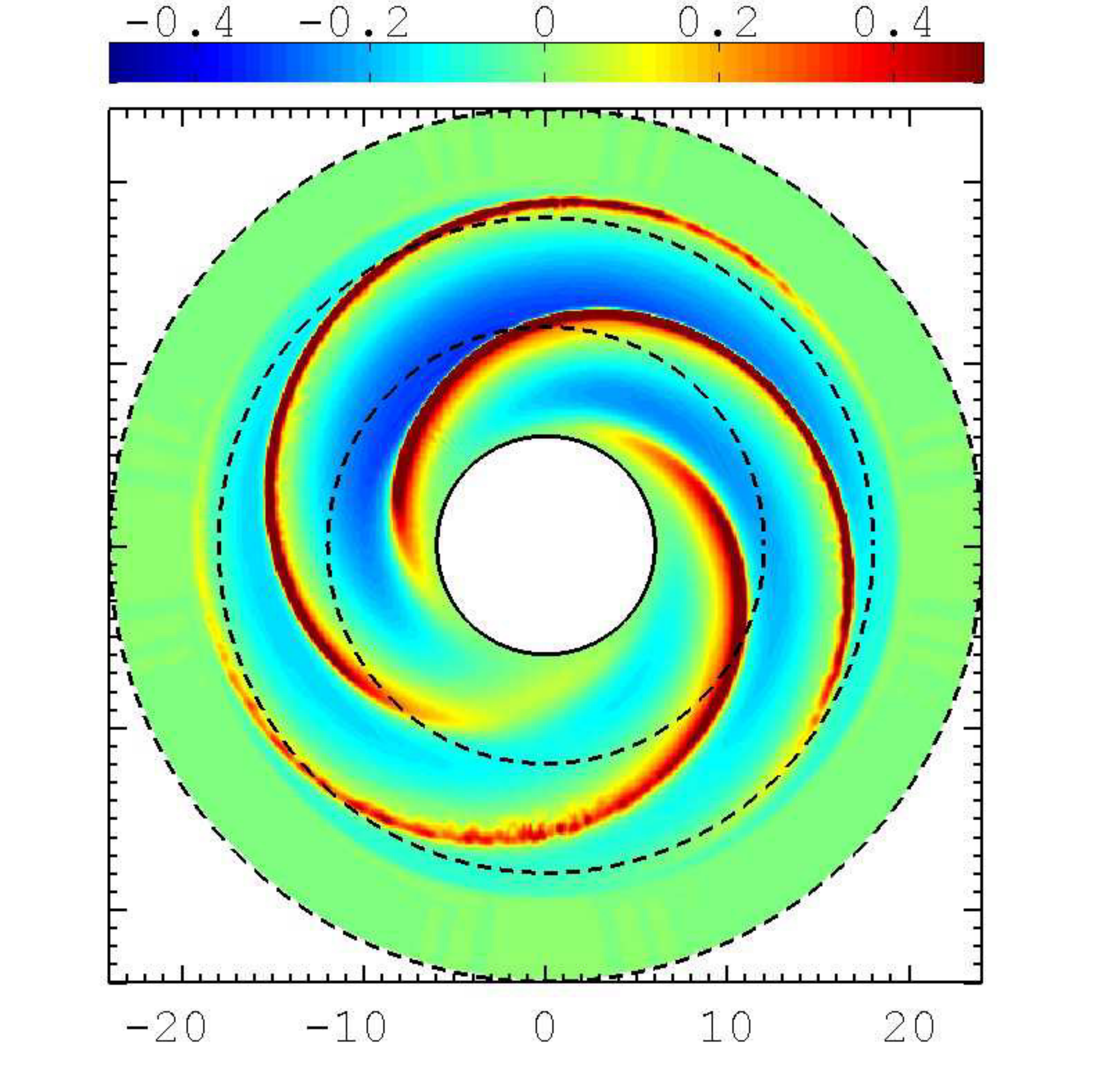}
\includegraphics[width=0.24\hsize]{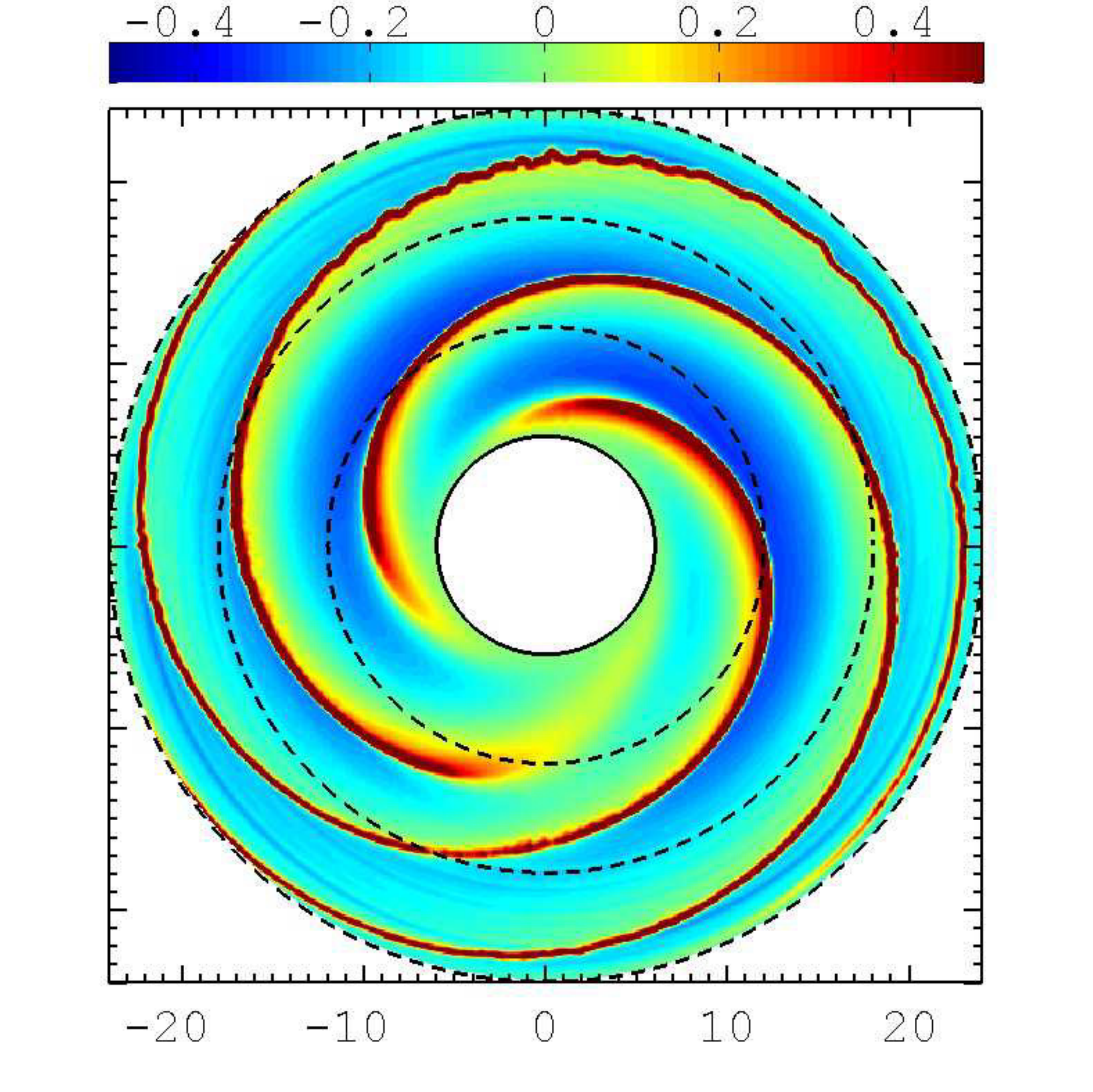}
\includegraphics[width=0.24\hsize]{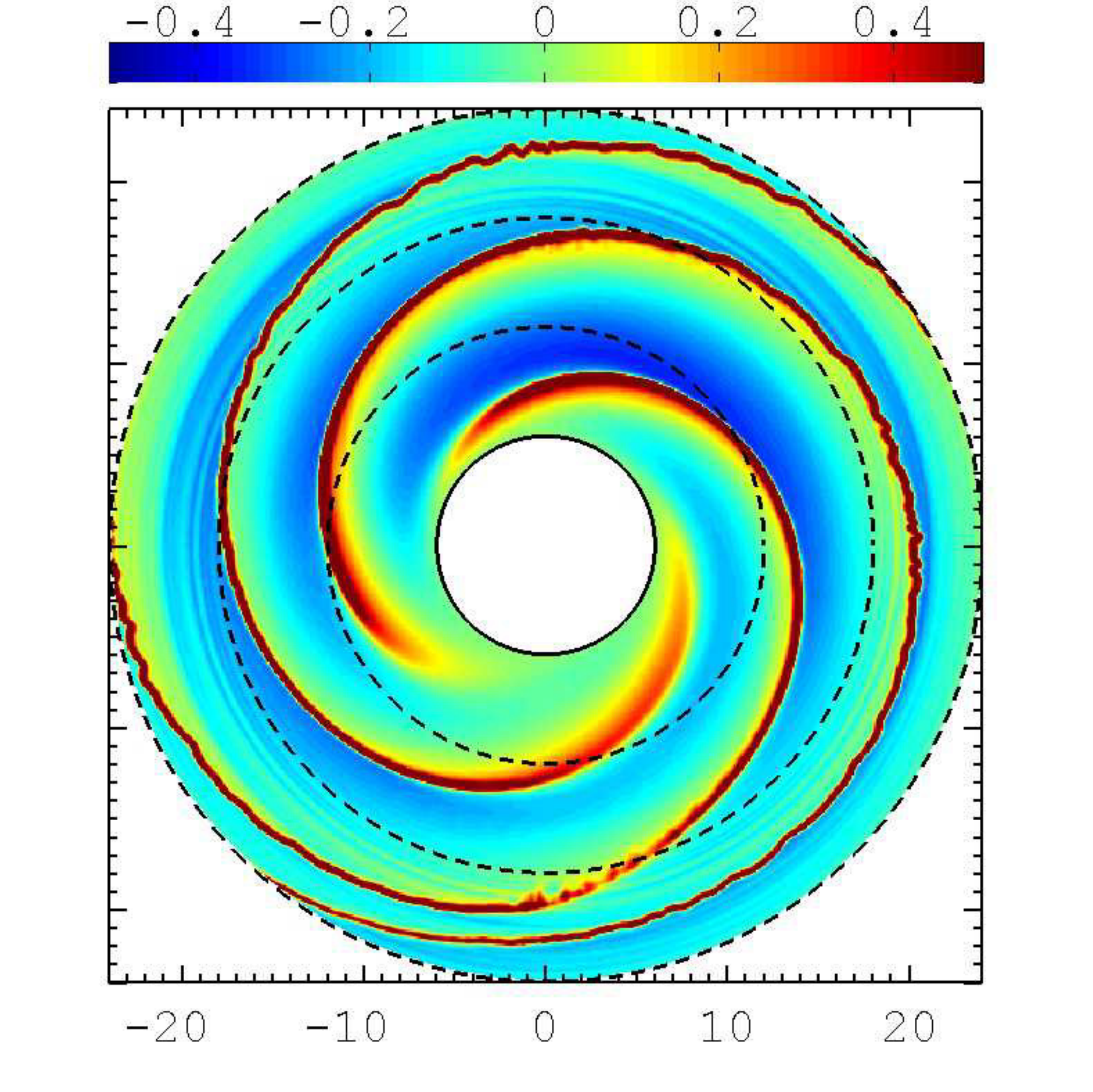}
\includegraphics[width=0.24\hsize]{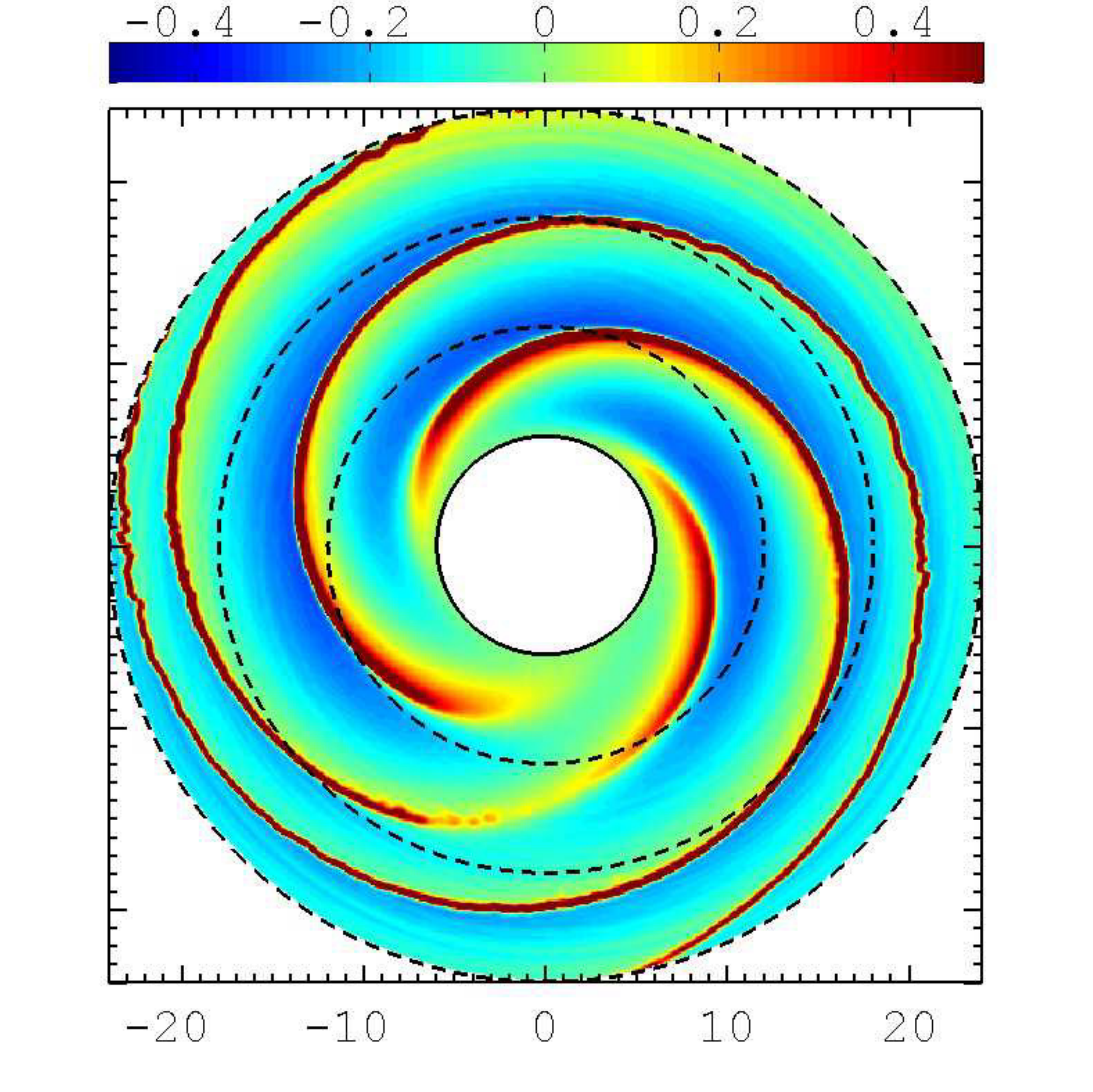}
\includegraphics[width=0.24\hsize]{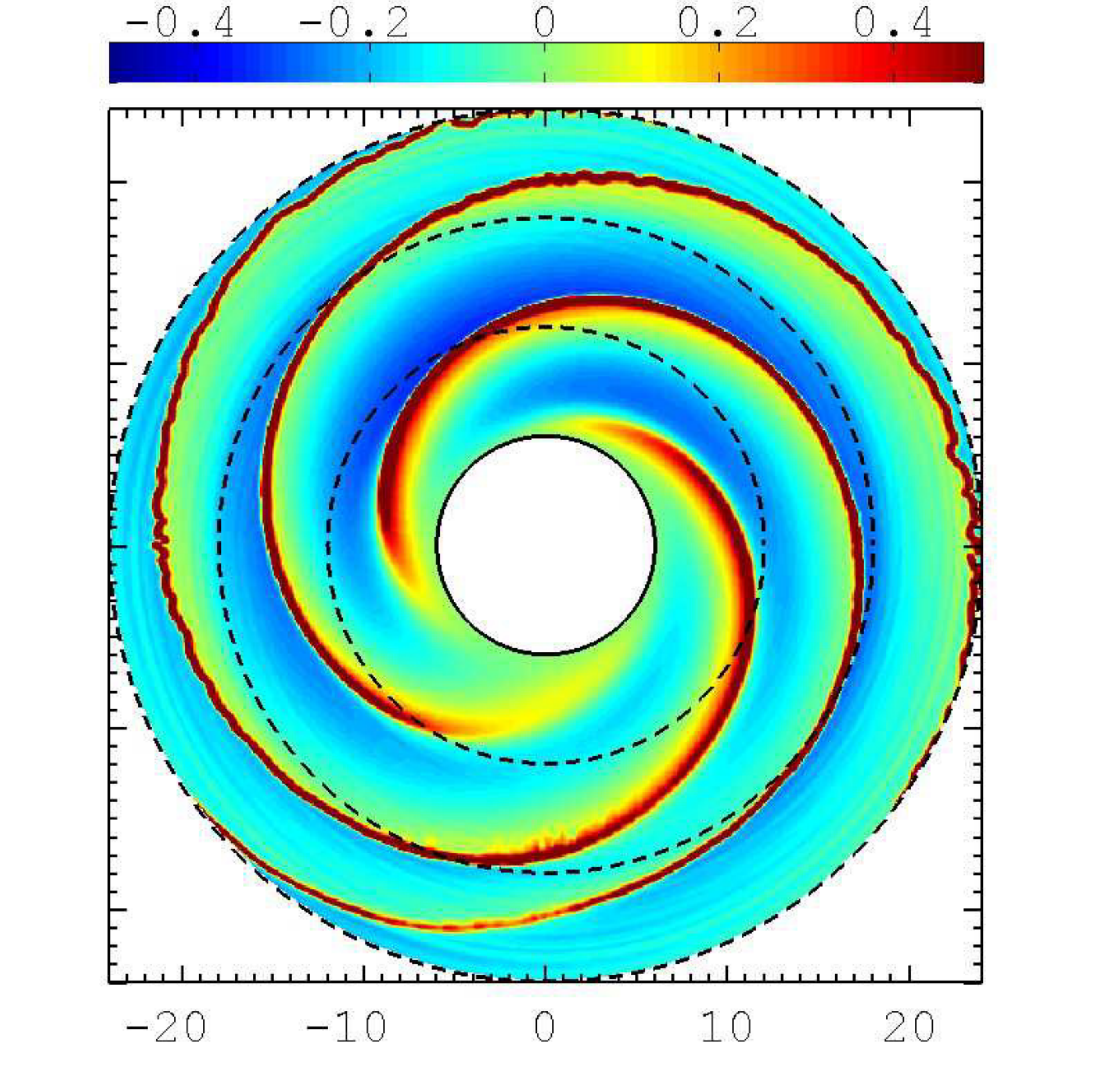}
\includegraphics[width=0.24\hsize]{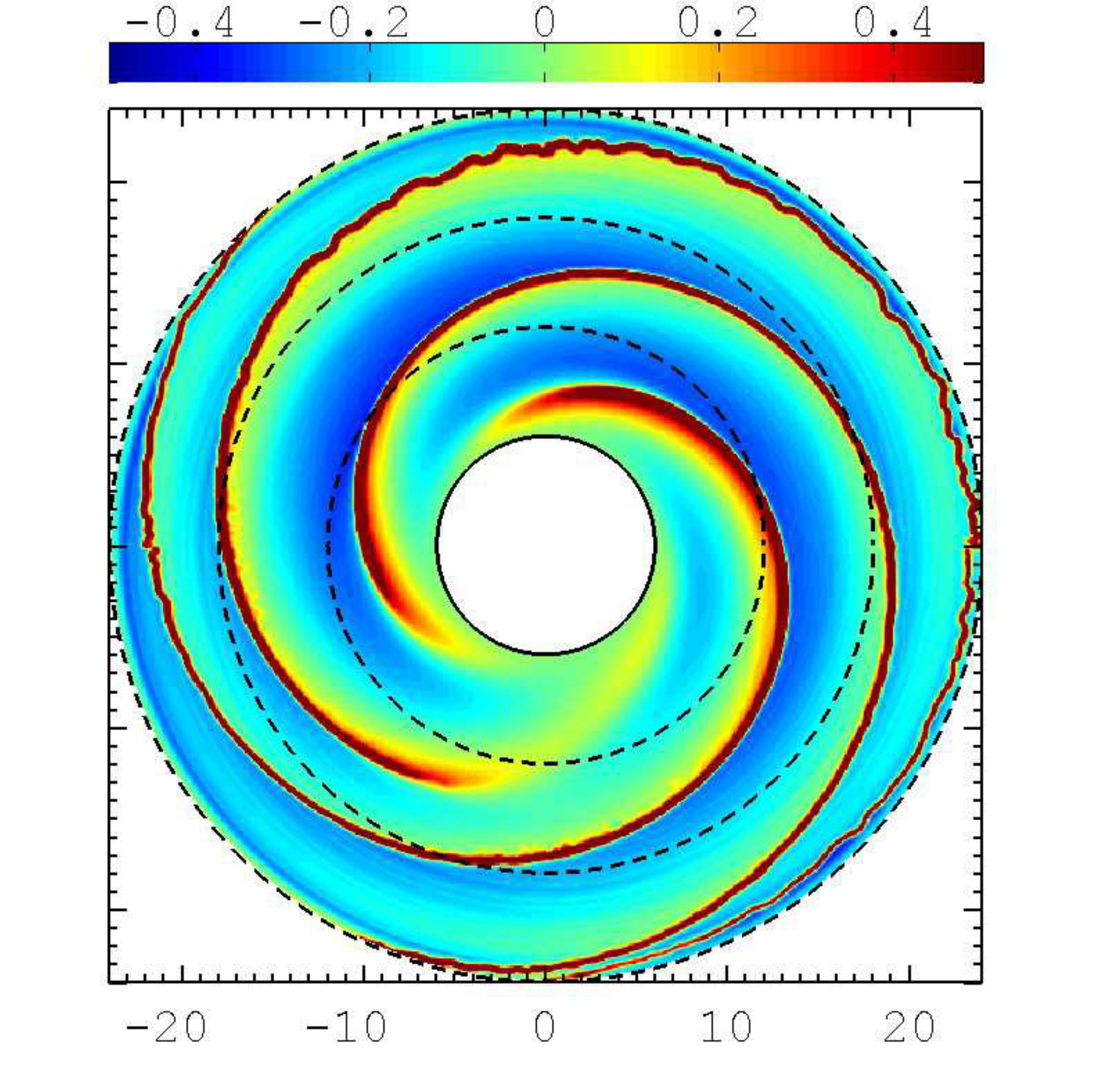}
\includegraphics[width=0.24\hsize]{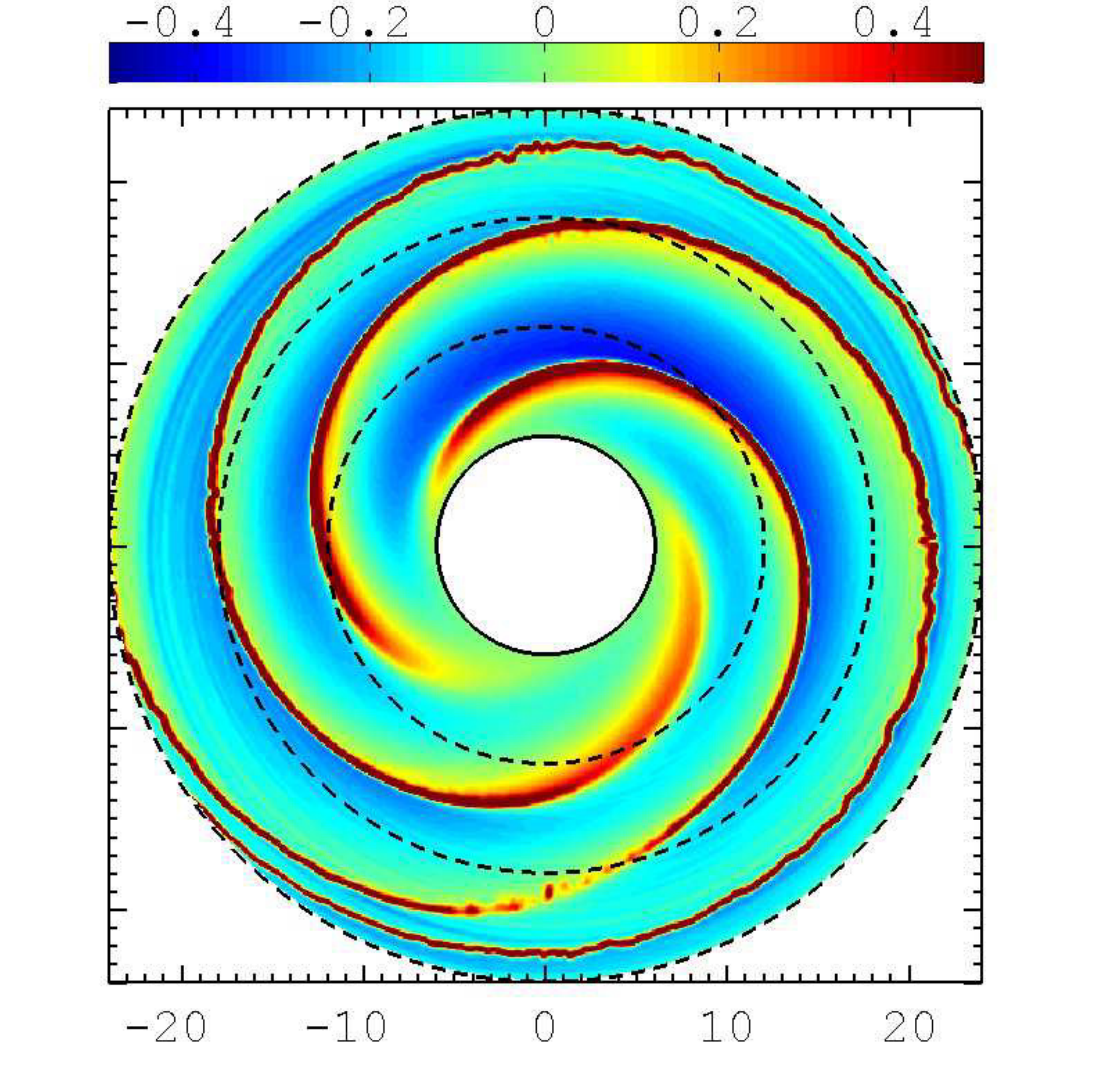}
\caption{Evolution of the relative surface density perturbation ($\Sigma_1/\langle\Sigma_g \rangle$) in model J1 with three modes imposed at the inner boundary, at times $t = 1.1; 1.4; 1.7; 2; 2.3; 2.6; 2.9; 3.2$ in units of $\approx 500$~Myr. Black circles are drawn at radii $6h$, $12h$, $18h$, and $24h$. The plots are drawn in the inertial, nonrotationg frame of reference.}\label{fig::evolution2}
\end{figure*}

\subsection{Resolution study}\label{sec::res_study}
\begin{figure*}
\includegraphics[width=0.33\hsize]{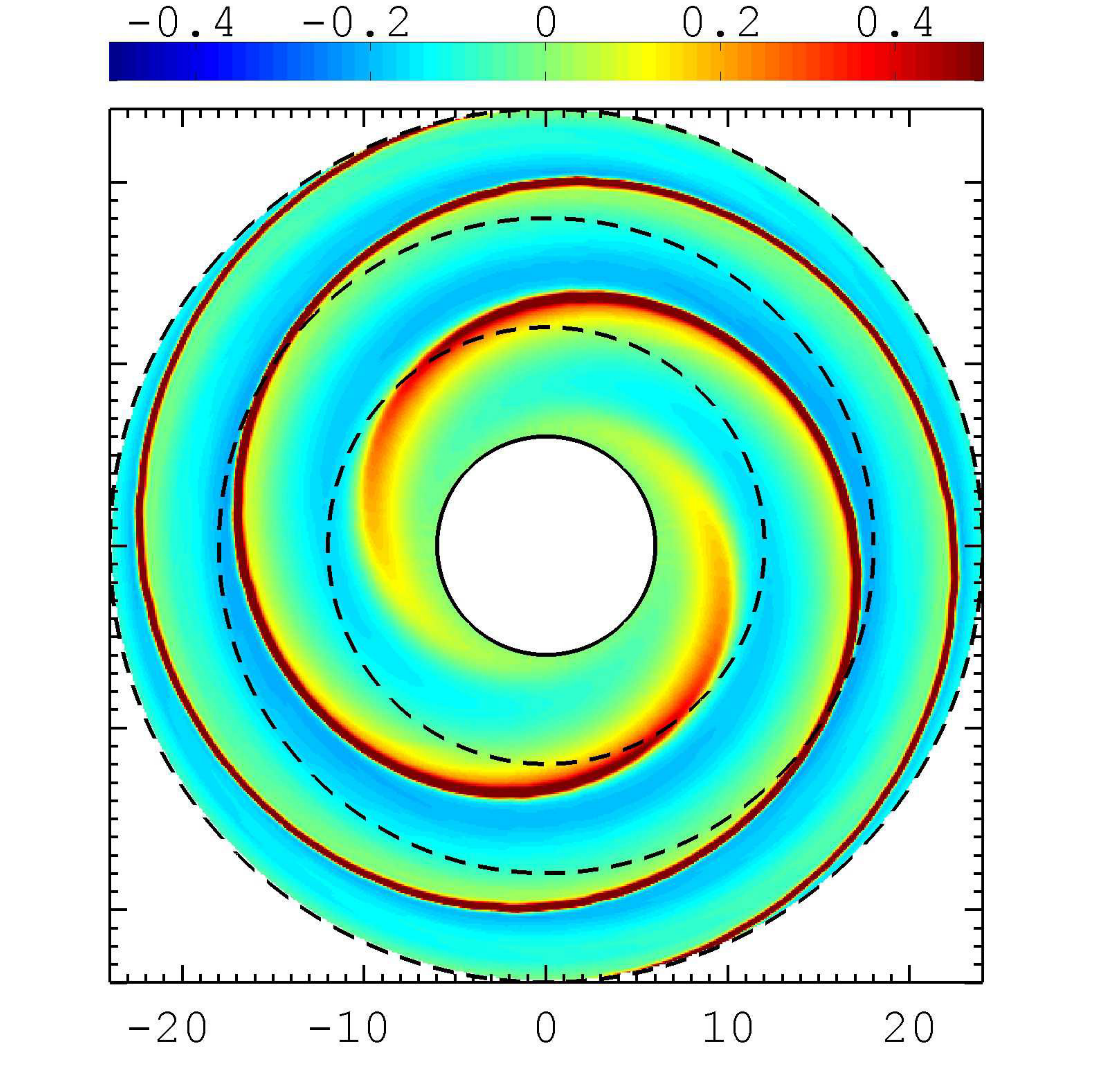}
\includegraphics[width=0.33\hsize]{B1_Sigma2D0018b.eps}
\includegraphics[width=0.33\hsize]{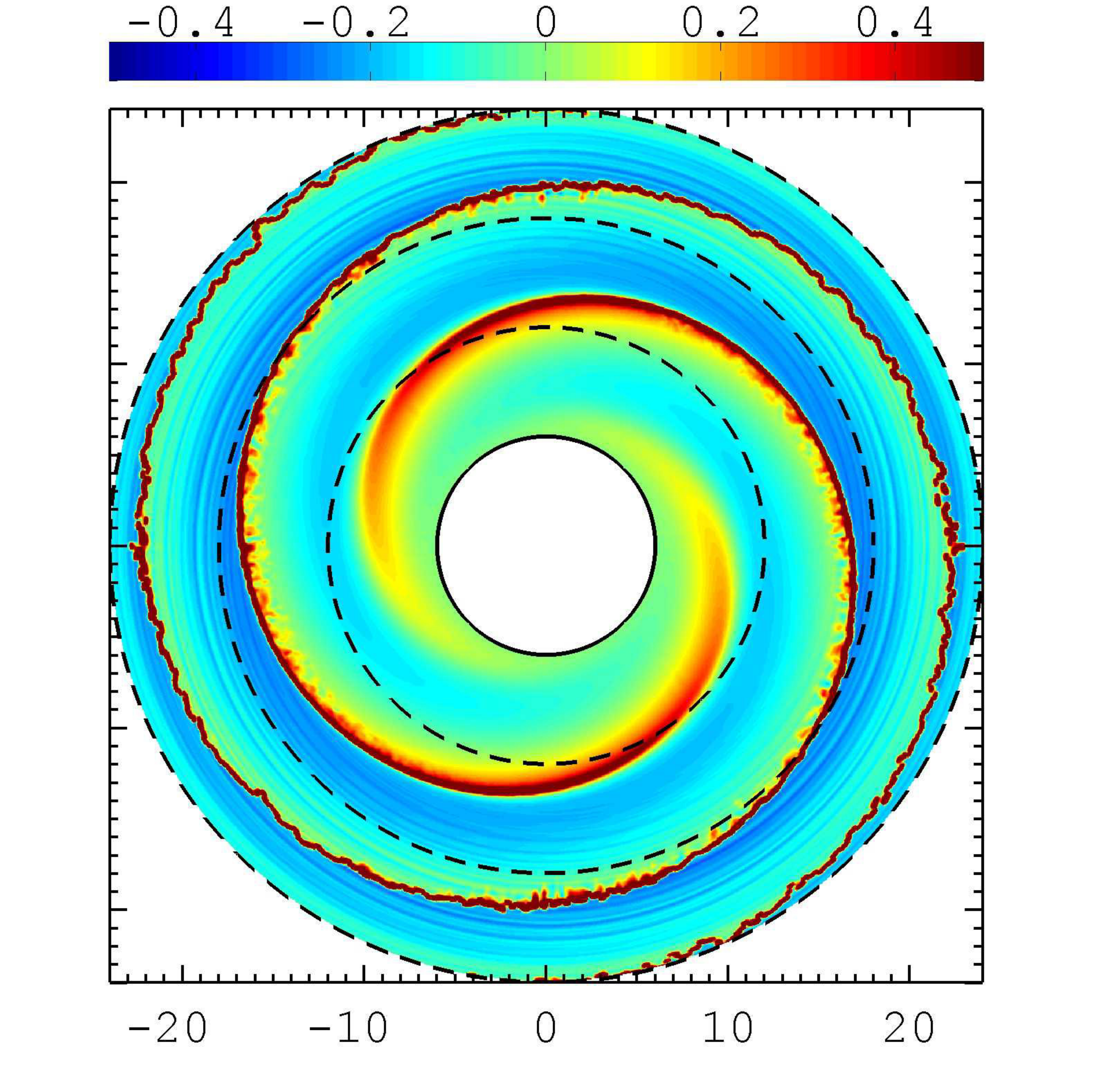}
\caption{Relative surface density perturbation ($\Sigma_1/\langle\Sigma_g \rangle$) at $t=2.5T_1$ for the B1 model simulated with varied spatial resolution, that is, different linear cell size: left --- 200~pc (low resolution), center --- 70~pc (reference case), right --- 35~pc (high resolution). Black circles mark different radii, at $6h$, $12h$, $18h$, and $24h$.}\label{fig::resolution}
\end{figure*}

All the models described so far have the same spatial resolution, that is, a cell size of $70$~pc in physical units. In order to check whether spatial resolution affects our general results, we have performed simulations with lower ($200$~pc) and higher ($35$~pc) cell linear size. Figure~\ref{fig::resolution} shows the results of these simulations. The general grand-design spiral structure is basically unchanged, as expected. Obviously, the narrow shock is smoother in the low resolution simulation;  higher spatial resolution let us resolve instabilities growing on very small scales related to shear flows behind the shock~\cite[see also][]{2004MNRAS.349..270W}.

In conclusion, our reference $70$~pc resolution is likely to be sufficient for studies of the global spiral structure and the detection of the shock instability. However, as to issues  related to local gravitational instabilities and star formation processes in the outermost gaseous discs in galaxies, more detailed simulations with higher spatial resolution are desired.

\section{Discussion and Conclusions}

In this paper we have presented a set of 3D~hydrodynamical simulations that describe the establishment of large-scale regular spiral structure in the outermost gaseous disc of spiral galaxies. The general scenario considers a galaxy in which spiral modes are excited inside the bright optical disc through the transfer of angular momentum to the outer regions by means of short trailing density waves. In the gas, such outgoing density waves can leak through the outer Lindblad resonance and propagate outwards  even when the gas effective velocity dispersion is above the condition of marginal stability with respect to axisymmetric perturbations (or, correspondingly, the column density is below the related critical density). In the simulations of the outermost gaseous disc, these outgoing waves are imposed as a stationary disturbance at the inner boundary. We have investigated the case in which galaxies and the corresponding boundary conditions are dominated by a single mode and, separately, the case in which more than one important mode is present. The results that we have obtained can be summarized as follows:

(i) The simulations are run as studies of a time-evolving situation in which the inner boundary acts as a source of density waves. After a relatively rapid initial transient, a quasi-stationary spiral structure is established over the entire outer disc, well outside the bright optical disc.

(ii) The simulations exhibit very good quantitative agreement with the predictions of the linear theory by~\cite{2010A&A...512A..17B} out to $r \approx 1.5 r_{opt}$. At larger radii the amplitude of spiral structure increases beyond the reach of the linear theory and then saturates. In this sense, our model is reminiscent of nonlinear tsunami-like waves. Correspondingly, spiral shocks form in the outermost regions, as a result of the supersonic motion of the patterns through the gaseous medium. Spiral shocks tend to be Kelvin-Helmholtz unstable in the post-shock regions. 

(iii) The simulations suggest that the outer spiral structure may be associated with significant star formation. As in a galactic tsunami, small amplitude perturbations become stronger and stronger at large radii, so that the shocks formed might trigger star formation events; some indications of star formation are indeed noted in UV observations. We note that the process studied in this paper might explain an outer UV star-forming ring in isolated galaxies even in the absence of an impact by an external object~\citep{2014MNRAS.439..334I}.   In our simulations, we did not investigate the star formation processes and the issue of the expected UV flux in great detail. The main reason for this is that to carry out a proper investigation of these processes would require a deep study of the conditions of the gaseous medium in the galactic outskirts, which are at present not well constrained by the observations; furthermore, we should have dealt with issues related to the resulting IMF, which are even less known. In this respect, we think that producing from the simulations synthetic UV spectra to be compared with the observations would be premature. In turn, in our simulations we focused on the larger-scale dynamical aspects of the processes involved during the propagation of density waves in the outermost gas disc.

(iv) The simulations of more realistic models, including a variety of physical factors, exhibit a generally similar behaviour in relation to the large-scale spiral structure. This suggests that the results obtained are rather robust and that indeed prominent large-scale spiral patterns should be a natural feature of galaxies with a gaseous disc extending beyond the optical radius. The simulations by~\citet{2008ApJ...683L..13B}~\citep[see also][]{2010ApJ...713L..780B} are interesting and exhibit some features in common with the results of our paper (in particular, the frequent finding of organized compression regions and filamentary structures). However, we wish to note that the study by~\citet{2008ApJ...683L..13B}: (1) focuses on a "fiducial star formation law"; (2) pays little attention to the relation between spiral structure in the bright optical disc and spiral structure in the outer gaseous disc; (3) appears to support the picture that star formation is expected only if the gas layer is close to conditions of local Jeans instability; (4) appears to depend on the addition of an extended outer disc with a {\it constant density}; (5) does not discuss the role of the thickness of the gaseous layer (which is a crucial factor in determining its stability). 

 (v)  In the picture proposed in this paper, the level of structuring of star formation regions in the outermost disc should reflect the level of regularity of the spiral structure in the bright optical disc. In other words, the spiral structure observed in the gas outside the bright optical disc should be characterized by well-organized, structured spiral patterns and filamentary structures if the bright optical disc is dominated by a grand-design structure. In contrast, the outermost spiral arms are expected to be less structured if the bright inner disc is dominated by many modes. This is a prediction that could be tested by studying a sufficiently large sample of spiral galaxies with extended gaseous discs. Of course, such a study would go well beyond the scope of this paper. In addition, we note that our simulations suggest clearly the possibility of ring-like star-formation regions, because we showed that the pitch angle of spiral patterns tend to decrease with radius; evidence of the relation between these morphological aspects and the underlying flows is likely to show up as brighter HI and, possibly, UV emission.

Of course, there are a few other interesting issues that are beyond the scope of this paper and await further investigations. One of these issues is the possible use of the observed spiral structure in the context of this paper to diagnose the amount and distribution of dark matter in the outer regions. To this purpose, this study should be further validated by detailed comparisons with observations in individual objects.

\section{ACKNOWLEDGMENTS}
We wish to thank the Referee for interesting suggestions and comments. The numerical simulations have been performed at the Research Computing Center~(Moscow State University) under grant 14-22-00041 and Joint Supercomputer Center~(Russian Academy of Sciences). This work was partially supported by the President of the RF grant (MK-4536.2015.2), RFBR grant (15-32-21062) and by the Italian MIUR.  SAK has been supported by a postdoctoral fellowship sponsored by the Italian MIUR.

\bibliography{outer_arxiv}

\end{document}